\documentclass{article}

\usepackage{epsfig}
\usepackage{amsmath}
\usepackage{amssymb}

\newtheorem{assumption}{Assumption}
\newtheorem{claim}{Claim}

\renewcommand{\leq}{\leqslant}
\renewcommand{\geq}{\geqslant}

\newcommand{\dif}{\mathrm{d}}
\newcommand{\erf}{\mathrm{erf}}

\begin{document}

\title{
    Compact Routing on Internet-like Graphs
}

\author{
    Dmitri Krioukov \thanks{\tt dima@krioukov.net} \and
    Kevin Fall \thanks{\tt kfall@intel-research.net} \and
    Xiaowei Yang \thanks{\tt yxw@mit.edu}
}

\maketitle

\begin{abstract}

\noindent The Thorup-Zwick (TZ) routing scheme is the first
generic stretch-3 routing scheme delivering a nearly optimal local
memory upper bound. Using both direct analysis and simulation, we
calculate the stretch distribution of this routing scheme on
random graphs with power-law node degree distributions, $P_k \sim
k^{-\gamma}$. We find that the average stretch is very low and
virtually independent of $\gamma$. In particular, for the Internet
interdomain graph, $\gamma \sim 2.1$, the average stretch is
around $1.1$, with up to 70\% of paths being shortest. As the
network grows, the average stretch slowly decreases. The routing
table is very small, too. It is well below its upper bounds, and
its size is around $50$ records for $10^4$-node networks.
Furthermore, we find that both the average shortest path length
(i.e.~distance) $\overline{d}$ and width of the distance
distribution $\sigma$ observed in the real Internet inter-AS graph
have values that are very close to the minimums of the average
stretch in the $\overline{d}$- and $\sigma$-directions. This leads
us to the discovery of a unique critical quasi-stationary point of
the average TZ stretch as a function of $\overline{d}$ and
$\sigma$. The Internet distance distribution is located in a close
neighborhood of this point. This observation suggests the
analytical structure of the average stretch function may be an
indirect indicator of some hidden optimization criteria
influencing the Internet's interdomain topology evolution.
\end{abstract}

\section{Introduction}

The recent observations of and scalability concerns with the
dynamics of the BGP routing table size growth,
\cite{huston-bgp-site,huston01-01,huston01-03,huston01-02,routing-requirements,BroNeCla02},
bring up the question of how small the routing table sizes for
distributed routing on realistic massive graphs can be made {\em
in principle}. In other words, what are the {\em fundamental\/}
limits of compactness of graph-theoretic routing in such networks?

Answering this question involves two things. On one hand, it calls
for assessment of results obtained in the area of distributed
routing. Since our first interest is the lower limits that can be
achieved {\em in principle}, we are more concerned with idealized
{\em static\/} routing in this paper. Being the simplest and most
fundamental routing model, static routing is where such limits can
manifest themselves. In the more complicated dynamic case, these
limits can only be higher.

On the other hand, answering the above question also requires
understanding of the basic properties of {\em realistic\/} massive
growing networks, the Internet being a good example of those. The
structure and evolution of these networks are subjects of
intensive studies these days.

\subsection{Previous work}

The above two research fields virtually do not intersect. We are
not aware of any routing schemes designed specifically for
scale-free graphs, and, vice versa, the literature concerned with
the properties of scale-free nets has not addressed the routing
problem yet. We tend to explain this lack of overlapping by the
fact that understanding of the nature of large networks observed
in reality started just recently. We review the results relevant
to our work in the two separate subsections.

\subsubsection{Routing}\label{sec:routing}

Since the pioneering work by Kleinrock and Kamoun, \cite{KK77},
the trade-off between stretch\footnote{The stretch factor is a
(usually worst-case) ratio of the path length produced by a
routing scheme to the shortest path length. Stretch-1 routing and
shortest path routing are synonymous.} and the amount of routing
information (i.e.\ memory space) required by a routing scheme has
been understood, analyzed, and improved. Many relatively recent
``Internet routing architecture'' proposals, \cite{nimrod,islay},
are based on the ideas of \cite{KK77}. Since \cite{KK77} was the
first work of its type, it is not surprising that na\"{\i}ve
routing based on it would generate stretch that is far from
optimal. Indeed, the simple calculations presented in Appendix
\ref{sec:kk_stretch} show that the stretch produced by \cite{KK77}
on the present Internet interdomain graph would be of the order of
10.

However, the high value of stretch was not the central problem
with \cite{KK77}. This work along with the works that closely
followed it, \cite{KK80,perlman85}, concentrated on optimal
hierarchical network clustering (splitting into areas) satisfying
a set of assumptions, but no proof of existence of such clustering
for generic graphs and no algorithm to find it when it exists were
obtained. While several subsequent works tried to overcome these
problems, all known hierarchical routing schemes eventually turned
out to be inferior with respect to more recent {\em direct\/}
routing schemes.\footnote{A routing scheme is {\em direct\/} if
the output port calculated at every node depends on the
destination label and nothing else. This implies that message
headers cannot be altered by intermediate nodes, and, hence, paths
produced by a direct routing scheme cannot have loops, which
justifies word ``direct.'' Many hierarchical routing schemes are
not direct.}

In the work by Peleg and Upfal, \cite{PeUp89}, the trade-off
between memory space and stretch for {\em generic}\footnote{That
is, {\em all}. A generic routing scheme is applicable to all
graphs.} networks was rigourously analyzed for the first time. The
work contained several issues that were addressed in many
publications that followed \cite{PeUp89}. One of the issues was
that only the total (per network) space was bounded, the local
(per node) space was unbounded. Another issue was that the scheme
required relabelling of nodes.

The fact that useful information about network topology can be
embedded in node labels to reduce the space is most easily seen in
the case of ring networks, where the local lower bound for
shortest path routing is $\Omega(n)$ if relabelling is not
allowed, and $\Omega(1)$ if nodes on the ring can be labelled
sequentially.

The most efficient label set for an $n$-node network is obviously
$[1,n]$.\footnote{The $[1,n]$ label set is relevant to a specific
terminology used in the routing literature. The common term
``routing table'' denotes a direct routing scheme with labels from
this set.} The fundamental lower bound is obtained in
\cite{EiGaPe03}. It is shown there that if the label space is
$[1,n]$, then, for {\em any\/} stretch (including shortest path
routing), there cannot exist a loop-free generic routing scheme
that would guarantee the local space less then $\sim3.7n^{1/2}$.

For {\em shortest path\/} routing, the lower bound turns out to be
higher. The pessimistic but intuitively expected results obtained
in \cite{GaPe96} show that for {\em any\/} stretch-1 routing
scheme, there exists a graph with maximum node degree $d$,
$\forall d\in[3,n)$, such that $\Omega(n\log d)$ bits of memory
are required at $\Theta(n)$ nodes. Since the trivial upper bound
for shortest path routing\footnote{The outgoing port is listed for
every destination.} is also $O(n\log d)$, \cite{GaPe96}
effectively demonstrates {\em incompressibility\/} of generic
shortest path routing.

Fortunately, the majority of graphs are slightly better. Applying
the Kolmogorov complexity theory to routing, the authors of
\cite{BuHoVi99} provide many upper and lower bounds for {\em
almost all\/} graphs. In particular, for shortest path routing,
not more than $3n$ (but not less than $n/2$) bits per node are
shown to be enough for the $1-1/n^3$ portion of all
$[1,n]$-labelled graphs. If labelling is relaxed to allow for
$O(\log^2 n)$-sized labels, then the local memory space upper
bound is reduced to $O(\log^2 n)$. The question about the space
lower bound for stretch-1 routing on almost all graphs with free
relabelling remains open.

While the results from the complexity theory or classical random
graph theory (\cite{FlaLeMa98,GaPe98}) for almost all graphs may
induce some optimism, very little can be said, on the practical
side, about if all the graphs from a given class of graphs are
good or bad with respect to routing compactness. Exploring
specifics of various graph families, a number of compact routing
schemes for special types of graphs have been constructed. There
are several results for rings, complete networks, trees, grids,
decomposable, planar\footnote{A planar graph can be drawn on a
plane without crossing edges.}, outplanar\footnote{An outplanar
graph can be embedded in a plane with all its vertices lying on a
convex polygon.}, bounded genus\footnote{The genus of a graph is
the minimum number of edge crossings with which a graph can be
drawn on a plane. Planar graphs have genus 0.},
chordal\footnote{The graph is chordal if all its cycles longer
than 4 have chords, or, equivalently, if it does not have induced
cycles longer than 3.}, etc., graphs,---but none so far for
scale-free graphs (even in the ``almost all'' context). It is easy
to explain since the properties of scale-free networks from the
graph-theoretic perspective are not fully understood yet.

Thus, the only currently existing tool to analyze the limits of
compactness of routing on scale-free networks is generic routing
schemes. Generic shortest path routing is incompressible, which
means that if the memory space is to be reduced, then the stretch
must be increased.

The memory space lower bound dependence on stretch is not
``continuous.'' As shown in \cite{GaPe96}, any generic routing
scheme with the maximum stretch strictly less than $1.4$ must use
at least $\Omega(n\log n)$ bits of memory on some nodes of some
graphs. In other words, the lower bound for generic schemes with
stretch $s<1.4$ is the same as in the incompressible case of
shortest path routing (consider the bound of $\Omega(n\log d)$
discussed above and take $d=\Theta(n)$). Furthermore, as shown in
\cite{GaGe01}, the lower bound for schemes with stretch strictly
less than $3$ is nearly the same as for shortest path
routing---$\Omega(n)$ bits of memory on some nodes of some graphs.

The minimum stretch factor that allows for significant memory
space lower bound decrease is 3. Cowen introduces a very simple
direct stretch-3 routing scheme with the local memory space upper
bound of $O(n^{2/3}\log^{4/3}n)$ in \cite{cowen01}. The scheme
uses relabelling, and labels are of size $3\log n$. In
\cite{ThoZwi01b}, Thorup and Zwick improve Cowen's result and
deliver a local space upper bound of $O(n^{1/2}\log^{1/2}n)$. They
also show how to implement routing decisions at {\em constant\/}
time per node and to reduce the label size to $(1+o(1))\log n$. We
call the above two stretch-3 schemes the {\em Cowen\/} and {\em
Thorup-Zwick\/} (TZ) schemes respectively.

The local memory space upper bound provided by the TZ scheme is
nearly (up to a logarithmic factor) optimal (best possible) since,
as demonstrated in \cite{ThoZwi01a}, any generic routing scheme
with stretch strictly less than $5$ must use at least
$\Omega(n^{1/2})$ bits of memory on some nodes of some graphs (see
Table \ref{table:stretch_1-3-5}). To the best of our knowledge,
the TZ scheme is the single generic stretch-3 routing scheme
delivering a nearly optimal local memory upper bound today. This
makes it ``exceptional'' in a sense that it delivers a nearly
optimal first possibility to decrease the local space down from
the shortest path routing incompressible limits. This also
explains why it is a primary subject of our present work.

\begin{table}
  \centering
  \caption{Total space lower bounds for low stretch values.}
  \label{table:stretch_1-3-5}
  \begin{tabular}{|c|c|c|}
    \hline
    Stretch $s$     & Lower bound           & Source \\
    \hline\hline
    $1 \le s < 1.4$ & $\Omega(n^2 \log n)$  & \cite{GaPe96} \\
    \hline
    $1.4 \le s < 3$ & $\Omega(n^2)$         & \cite{GaGe01} \\
    \hline
    $3 \le s < 5$   & $\Omega(n)$           & \cite{ThoZwi01a} \\
    \hline
  \end{tabular}
\end{table}

To finish the introduction to relevant results in routing, we
briefly touch on two more areas.

First, nothing confines one to considering only a multiplicative
stretch factor. The concepts of {\em additive\/} stretch---that
is, the additive error factor in distance approximation---and even
of mixed multiplicative-additive stretch have been introduced and
studied to some degree (\cite{AinCheInMo99,DorHalZwi00,ElPe01}).
Additive stretch models are potentially better suited for graphs
with low average distances (the Internet interdomain graph, for
example) since, as can be easily seen, the {\em short\/} distances
are harder to approximate than the long ones
(\cite{AinCheInMo99,cowen01}\footnote{In fact, one of the very
important component of many routing schemes including the Cowen
scheme is a careful balance between short and long paths.}).
However, there are very few {\em routing schemes\/} based on
additive stretch. The latest one is probably a very efficient
additive stretch-2 routing scheme for chordal graphs by
Dourisboure and Gavoille, \cite{DouGa02}.

Second, the above discussion concerns the static case only. In
case of dynamic networks, the other components that add complexity
to the picture are stability issues as well as adaptation or
communication costs\footnote{That is, the number of messages
generated per topology change, their sizes, or the total amount of
data sent to guarantee communication, etc.}.

One of the first works that rigourously addresses the problem of
routing in dynamic networks is \cite{AfGaRi89}. The ``positive''
result in the paper is a dynamic routing scheme for growing trees.
One of the interesting ``negative'' results is the analysis of the
trade-off between the stretch and adaptation cost. Assuming that
the space and message sizes are unbounded, \cite{AfGaRi89} shows
that any generic dynamic routing scheme with stretch $s<k$ must
send at least $\Omega(n/k)$ messages per topology change on some
networks.

In a recent work by Krizanc, Luccio, and Raman, \cite{KriLuRa02},
three schemes for dynamic routing on rings with different
stretch-space-adaptation trade-offs are constructed. All of the
three schemes are dynamic versions of interval routing
schemes.\footnote{A review of interval routing is given in
\cite{gavoille00}.}

The {\em Bubbles\/} model, \cite{DoKraKriPe99}, is a {\em
generic\/} dynamic routing scheme that uses hierarchical
partitioning of the spanning tree of a graph. Because of the
specifics of the network model considered in \cite{DoKraKriPe99},
the stretch factor is replaced there by the ``super-hop count,''
the maximum number of hops produced by the scheme, and the
adaptation cost is measured by ``adaptability,'' the maximum
number of nodes affected by topology change updates. The most
efficient variant of the scheme, designed for high-degree
networks, provides the local memory space upper bound of
$O(kn^{1+1/k}\log d)$, where $k$ is the maximum number of hops and
$d$ is the maximum node degree. The adaptation cost upper bound is
$O(3^kn^{1/k}d)$ with node failures and $O(3^kn^{1/k})$ with
link-only failures. The adaptation cost lower bound for low-degree
graphs ($d=O(1)$) is also obtained. It is $\Omega(n^{1/k})$.

Finally, in the context of the {\em end-to-end communication\/}
problem (see review by Fich~\cite{fich98}), the stretch factor is
not explicitly considered---the problem is just to guarantee
communication between two fixed nodes in presence of frequent
network failures and to optimize the trade-off between the total
number of messages generated per data item (communication cost)
and required local memory space per incident link. For the first
solution that is polynomial in communication cost and logarithmic
in memory space, see the work by Kushilevitz, Ostrovsky, and
Ros\'{e}n~\cite{KuORo98}. For the latest memoryless network
result, see~\cite{FraGa03}.

For more details on progress in routing, see the excellent review
by Gavoille~\cite{gavoille01} and the monograph by
Peleg~\cite{peleg01}.

\subsubsection{Scale-free networks}\label{sec:scale-free_networks}

Since the discovery of {\em power-law\/} distributions in the
Internet in \cite{FaFaFa99}, the Internet {\em scale-free\/}
nature has been a subject of very intense studies and generated an
enormous number of publications. A particularly interesting fact
is that the Internet appears to be just one example of scale-free
networks that have been found extremely ubiquitous. The list of
networks, within which power-law or, more generally, {\em
fat-tailed\/} distributions have been observed, include---beyond
the Internet, both at the interdomain and router levels
(\cite{VaPaVe02,SiFaFaFa03})---the WWW (\cite{AlJeBa99}), power
grids (\cite{AmScaBaSta00}), airport networks
(\cite{AmScaBaSta00}), biological (\cite{JeToAlOlBa00}),
ecological (\cite{MonSol00}), language (\cite{DorMen01a}), and
social (\cite{NewWatStr02}) networks, the latter including
scientific collaboration (\cite{newman01}), movie actor
collaboration (\cite{AmScaBaSta00}), human acquaintance
(\cite{AmScaBaSta00}), and sexual contact (\cite{LiEdAmStAb01})
networks.\footnote{See also \cite{WatStr98,watts99,newman03} and
\cite{nd-networks}.}

All the networks listed above involve an element of randomness.
Classical Erd\H{o}s-R\'{e}ni random $n$-node graphs,
\cite{ErRe59}, have links between every pair of vertices with the
uniform probability $p$. The ensemble of such graphs is called
${\cal G}_{n,p}$. Their average node degree is $\overline{k}\sim
np$, the node degree distribution is the Poisson distribution with
exponentially small number of high-degree nodes,
$P_k\sim\overline{k}^k\mathrm{e}^{-\overline{k}}/k!$, and average
distance is $\overline{d}\sim\log n/\log\overline{k}$,
\cite{bollobas85}.

All the networks observed in reality defer drastically from the
${\cal G}_{n,p}$ graphs. One of the differences of particular
importance for this paper can be seen as certain inconsistency
between the average distance and average node degree predicted by
the Erd\H{o}s-R\'{e}ni model. In the real Internet interdomain
$1.1\times10^4$-node graph, for example, $\overline{k}\sim5.7$ and
$\overline{d}\sim3.6$, \cite{VaPaVe02}\footnote{For more detailed
measurements of the Internet topology, see \cite{BroCla01a}.},
while the ${\cal G}_{1.1\times10^4,5.2\times10^{-4}}$ graphs have
$\overline{d}\sim5.3$. The ${\cal G}_{n,p}$ graphs of the same
size with the right average distance $\overline{d}\sim3.6$ would
have to have the average degree $\overline{k}\sim14$. In
Sections~\ref{sec:gnp} and~\ref{sec:minimum}, we see how strongly
these slight differences affect the average stretch.

The simultaneously small values of the average distance and
average node degree necessarily imply a larger portion of
high-degree nodes than in the classical random graphs. In other
words, the node degree distribution must be fat-tailed. The
power-law, $P_k\sim k^{-\gamma}$, one of such fat-tailed
distributions, is what has been observed in many networks listed
above, exponent $\gamma$ ranging between $2$ and $3$. For the
Internet interdomain graph, $\gamma\sim2.1$,
\cite{FaFaFa99,VaPaVe02}.

Both the ${\cal G}_{n,p}$ graphs and graphs with fat-tailed degree
distributions are often said to possess the {\em small-world\/}
property, \cite{milgram67}, to emphasize that they have extremely
low average distances (for networks of such size), even though
average distances in ${\cal G}_{n,p}$ graphs are slightly higher.
The famous play {\em Six Degrees of Separation}, \cite{guare90},
is based on the observation made in \cite{milgram67} that the
average distance in human acquaintance networks is around~$6$. As
far as routing in the Internet is concerned, the simple but
critically important fact that the Internet distance distribution
has very low values of mean and dispersion (that is, that there
are virtually no remote points) gets fairly often either
overlooked or neglected.

Networks with fat-tailed degree distributions are also called {\em
scale-free\/} since their node degree distribution lacks any
characteristic scale, \cite{BarAlb99}, in contrast to the ${\cal
G}_{n,p}$ graphs with the narrow Poisson degree distribution
centered around the characteristic average value $\overline{k}\sim
np$.

The explanation of appearance of fat-tailed distributions in
realistic networks is obviously a very important problem. While a
large number of models generating power-laws have been suggested,
arguably none of them so far captures the underlying principles of
the Internet evolution and predicts the observed Internet topology
well enough.

The most popular model is the model for growing networks with
preferential attachment\footnote{The probability for a new-coming
node to attach to a target node already in the network is
proportional to the target node degree.} by Barab\'{a}si and
Albert (BA), \cite{BarAlb99}. The BA model is very simple, it does
not have external parameters, which makes it attractive for a
physicist, but, in its ``pure'' form, it predicts $\gamma=3$. The
model can be freely modified to produce other values of $\gamma$
and even scaling behaviors deviating from power-laws,
\cite{AlBa02}, but its applicability to the Internet evolution has
been extensively criticized in \cite{WilCriticality,WilRevisited}.
In particular, in \cite{WilCriticality}, it is noted that the BA
model and its derivatives are capable of reproducing what has been
already measured but they fail to predict correctly anything new
about the Internet topology, anything that has not been measured
yet. As such, the BA model is not {\em explanatory\/} but {\em
descriptive}, in the terminology of \cite{WilCriticality}.

One of the most interesting models for the Internet evolution is
analyzed in \cite{FaKoPa02}. It is shown there that a simultaneous
(trade-off) optimization of last-mile costs (geometrical distance)
and average hop distance from the network ``center'' (the first
node arrived) can lead to power-laws. In other words, power-laws
can be a result of optimization of the trade-off between the
physical link cost and average delay associated with the average
path length in hops---in data networks, every hop is a source of
queuing delay and packet loss. Although there is arguably no such
optimization intentionally (in a controlled manner) happening in
the real Internet, which is driven primarily by its economy
outlined, for example, in \cite{wbnorton} and modelled in
\cite{KriouVol03,ChaJamWil03}, the link, noted in \cite{FaKoPa02},
to the Mandelbrot language model, \cite{mandelbrot53}, maximizing
language efficiency and resulting in power-laws, deserves some
serious attention.

It is worth mentioning that the whole subject of the scale-free
nature of the Internet has been doubted in \cite{LaByCroXie03},
where it is shown that just traceroute-based measurement
techniques\footnote{These obviously include both standard
traceroutes and BGP table dumps.} may be solely responsible for
the Internet {\em appearing\/} scale-free, and it may belong to
the ${\cal G}_{n,p}$ class in reality. Indeed, it is easy to see
that the farther a measured node is from {\em all\/} the measuring
nodes (sources of traceroutes) on a graph, the smaller portion of
the total number of links incident to the measured node can be
detected by traceroutes. Although some possibility for the
Internet {\em router level\/} topology to be less fat-tailed does
exist, the results of analysis in \cite{LaByCroXie03} itself
essentially rule out any possibility for the Internet {\em
interdomain\/} topology to deviate strongly from the power-law if
one considers the very low value of the almost undoubtedly
measured average distance and the high enough numbers of measuring
points used in various Internet interdomain topology studies (ten
vantage points are used, for example, in \cite{SuAgReKa02}).

On the practical side, a very useful work is presented in
\cite{TaGoJaShWi02}. The authors distill a set of criteria
assigning a characteristic ``signature'' to any type of graphs.
These signatures are used then to compare the real Internet
topology with topologies produced by various Internet topology
generators and with topologies of several standard types of graphs
(trees, grids, complete graphs, classical random graphs, etc.).
Surprisingly enough, the structural Internet topology generators
trying to incorporate the perceived hierarchical structure of the
Internet in their algorithms are found inferior to the
degree-based generators ``blindly'' reproducing the observed
degree distribution. The simplest generator of this type is the
PLRG generator suggested in \cite{AiChLu00}\footnote{The
construction procedure is due to Molloy and Reed,
\cite{MolRee95,MolRee98}.} and analyzed (among other generators)
in \cite{TaGoJaShWi02}. It is also found in \cite{TaGoJaShWi02}
that the only type of standard graphs having the same signature as
the Internet is complete networks.

Another important observation made in \cite{TaGoJaShWi02} is about
the presence of correlation between the link value, defined as a
weighted number of shortest paths passing via the link, and the
lower degree of the nodes attached to the link. This observation
is consistent with the earlier measurements of the Internet
interdomain graph in \cite{VaPaVe02a} showing that the node
betweenness, defined as the total number of shortest paths passing
via a node, is linearly correlated with the node degree.

These measurements point to the ``self-establishing'' nature of
the Internet hierarchy, which is further revealed by the
observations of the power-law decay of the clustering coefficient
in \cite{VaPaVe02a,VaPaVe02}. The clustering coefficient $c_k$,
defined as the average ratio of the number of $3$-cycles involving
$k$-degree nodes to its maximum value $k(k-1)/2$, measures how
close an average $k$-degree node neighborhood is to a clique. Its
power-law decay for the Internet interdomain graph indicates that
small, low-degree ASs\footnote{For the measurements of correlation
between AS size and degree, see \cite{TaDoGoJaWiSh01}.} tend to
create numerous, highly clustered structures that are connected
with each other via a sparse formation of ``hubs''---large,
high-degree ASs.

It is interesting to note that the power-law decay of the
clustering coefficient has been observed only for
``uncontrolled,'' ``self-evolving'' networks---the Internet
interdomain graph, the WWW, biological, language, and social
networks. Networks with a stronger element of design and external
control---the Internet router level graph and power grids, for
example---do not exhibit such behavior,
\cite{VaPaVe02,RaBa03,RaSoMoOlBa02}. The clustering coefficient as
a function of node degree seems to be relatively constant in the
latter cases, while its average values are still much higher than
in the ${\cal G}_{n,p}$ graphs, which is another drastic
difference between the ${\cal G}_{n,p}$ model and real-world
networks, \cite{WatStr98,newman03}.

For the analytical part of our present work, we need to know the
distance distribution in scale-free graphs. The problem is very
hard and it has not been solved analytically yet. There are some
recent results on the {\em average\/} distance in scale-free
networks, \cite{ChLu03,CoHa03}. {\em Implicit\/} expressions for
the distance {\em distribution\/} are constructed in
\cite{CoDoHaKaMoSh02}. More {\em explicit\/} analysis of the
distance distribution is performed in
\cite{DoMeSa03a,FroFroHol02}.\footnote{Why the frequently referred
expressions derived in \cite{NewStrWat01} are imprecise is shown,
for example, in \cite{krzywicki01}.} Unfortunately, all these
results are valid only for static, equilibrium networks without
vertex-vertex degree correlations. All realistic scale-free
networks are growing, non-equilibrium. They necessarily have node
degree correlations resulting in much wider distance
distributions, \cite{dorogovtsev-private}. Surprisingly, the model
constructed by Dorogovtsev, Goltsev, and Mendes for deterministic
scale-free graphs in \cite{DoGoMe02} (the DGM model) turns out to
be capable of analytically producing a Gaussian distance
distribution similar to the distance distribution observed in the
real Internet,
\cite{VaPaVe02,VaPaVe02a,BroNeCla02,huston-bgp-site}. The width of
the Gaussian for a $10^4$-node network, $1.1$, is very close to
the width of the Internet interdomain distance distribution,
$0.9$, but the average distance is slightly higher---$4.8$ instead
of $3.6$. As noted in \cite{DoGoMe02}, simulation-based
measurements of the distance distribution in the BA model also
produce similar Gaussians, \cite{krzywicki01}.

For further details on scale-free networks, see the excellent
review \cite{DorMen02a} and book \cite{DorMen-book03} by
Dorogovtsev and Mendes.

\subsection{Our contribution}

One might expect that for scale-free graphs, the majority of known
generic routing schemes would be very inefficient. Indeed, many
routing schemes (including the Cowen and TZ schemes) incorporate
{\em locality} by carefully differentiating between close and
remote nodes. This approach makes routing more efficient (in the
stretch-versus-space trade-off sense) by keeping only approximate
(non-shortest path) routing information for remote nodes, while
full (shortest path) routing information is kept for local nodes.
In scale-free graphs characterized by low average distances and
distance distribution widths, local nodes comprise huge portions
of all the nodes in a network, so that one might suspect that
locality-sensitive approaches might break for such networks. For a
good example demonstrating that this might be quite plausible, see
the Appendix~\ref{sec:kk_stretch}, where the stretch factor is
found to be very high for the Kleinrock-Kamoun~(KK) routing
scheme~\cite{KK77} applied to the scale-free networks.

Furthermore, one can take the situation to its extreme and
consider a ``smallest-world'' graph, that is, a complete graph.
The idea is suggested in part by \cite{TaGoJaShWi02}, where the
Internet graph ``signature'' is found to be similar to the
complete network ``signature'' (cf.\ Section
\ref{sec:scale-free_networks}). On would find then that both the
Cowen and TZ average stretch factors in this extreme case of
complete graphs are high---as can be easily checked, the average
TZ stretch for a complete graph of size $n$ is
$2-n^{-1/2}\log^{-1/2}n-o(n^{-1/2})$.\footnote{The
Kleinrock-Kamoun average stretch is much worse, of course. It is
trivial to see that it grows as $\Theta(\log n)$.}

We find that the case of realistic scale-free networks with
Internet-like characteristics is significantly better.

We consider the TZ scheme, which is an ``exceptional'' routing
scheme in the sense explained in Section \ref{sec:routing}. Being
generic, the TZ scheme provides only general maximum stretch and
space bounds. It says nothing about the average stretch or stretch
distribution on a particular class of graphs.

We calculate, both analytically and via simulations, the TZ
stretch distribution on Internet-like topologies. The analytical
part of the problem is hard. It assumes knowledge of the distance
distribution in correlated scale-free networks. The exact form of
this distribution has not been obtained analytically yet (see
Section \ref{sec:scale-free_networks}). Given the observation that
the DGM model~\cite{DoGoMe02} analytically produces the Gaussian
distance distribution that is close to the real Internet distance
distribution, we choose to parameterize distance distributions in
small-world graphs we consider in this paper by Gaussian
distributions. To obtain our results we still have to make a
series of simplifying assumptions that are fully discussed in
Section~\ref{sec:analytical_results}.

For the simulation part, we develop our own TZ scheme simulator
and use it on graphs produced by our implementation of the PLRG
generator~\cite{AiChLu00}, the initial justifications for using it
being discussed in Section \ref{sec:scale-free_networks}. Since
the PLRG generator outputs uncorrelated networks, there are some
concerns regarding its capability of reproducing {\em all\/} the
features of strongly correlated nets, such as the Internet.
However, since, as we see in Section~\ref{sec:analytical_results},
the stretch distribution turns out to be a function of the
distance distribution and the graph size only, all we need from a
graph generator for our purposes is that distance distributions in
graphs produced by it be close to distance distributions observed
in real-world graphs. We find that PLRG-generated graphs with the
node degree distribution exponent $\gamma=2.1$ have the distance
distribution that is very close to the distance distribution
observed in the Internet.

We obtain a close match between the analysis and simulation data
for the average TZ stretch and stretch distribution in Section
\ref{sec:numerical_results_and_simulations}. We find that the
average stretch is {\em very low\/} and virtually {\em
independent\/} of exponent $\gamma$. In particular, in the case of
the Internet interdomain graph, $\gamma\sim2.1$ and size
$n\sim10^4$, the average stretch is $1.14$ and $1.09$ according to
the analysis and simulations respectively. The stretch
distribution has a peculiar form. The majority of paths produced
by the TZ scheme are shortest---up to 71\% according to the
simulations. The majority of non-shortest paths have stretch
values of $4/3$ and $5/4$. The portion of paths with other stretch
values is very small.

The average number of entries in the routing table\footnote{We are
still using term ``routing table'' here even though it is not
completely correct from the graph-theoretical perspective since
the TZ scheme does not use labels from the $[1,n]$ set.} is also
extremely low---well bellow its upper bounds. For graphs with the
Internet-like parameters, $n\sim10^4$, $\gamma\sim2.1$, it is
approximately $52$.

We also show that the average stretch slowly {\em decreases\/}
with the network growth even if the average distance scales as
$\log n$. However, the average stretch does not approach $1$ even
for sufficiently large $n$. Therefore, the amount of non-shortest
paths seems to be unbounded.

The fact that the average stretch on scale-free networks turns out
to be low is not by itself surprising. Indeed, a scale-free graph
can be described as a ``collection of interconnected stars,'' and
it is easy to see that both the Cowen and TZ average stretch
factors on stars are equal to 1. The average stretch is expected
to be substantially higher for other types of random networks. We
confirm this expectation in Section~\ref{sec:gnp} where we
calculate the average stretch for certain ${\cal G}_{n,p}$ graphs.

We also obtain a row of really surprising results presented in
Section \ref{sec:minimum}. The analytical expressions we provide
for the average stretch $\overline{s}$ allow us to consider it as
a function of the parameters of the distance distribution in a
graph, the parameters being the average distance $\overline{d}$
(the first moment) and distance distribution width $\sigma$ (the
square root of the second moment or the standard deviation).
First, we find that both $\overline{d}$ and $\sigma$ of the
distance distribution in the Internet are very close to the local
{\em minimums\/} of $\overline{s}(\overline{d},\sigma)$ in the
$\overline{d}$- and $\sigma$-directions respectively.

Next, simultaneous proximity of the Internet distance distribution
to the minimums of $\overline{s}(\overline{d},\sigma)$ in the both
directions makes us search for a stationary point\footnote{The
point is stationary if all first-order partial derivatives of a
function at this point are zero. This is a necessary condition for
an extremum. The function has a minimum (maximum) at a given
stationary point if all eigenvalues of the matrix of all
second-order partial derivatives of the function at this point are
positive (negative).} and potential extremum of $\overline{s}$.
Our analytical results allow us to collect enough data to
discover: 1) a region of $\overline{d}$ and $\sigma$, where
function $\overline{s}(\overline{d},\sigma)$ is concave and
stretch is particularly low, which we call the {\em minimal
stretch region\/} or the MSR; and 2) a unique critical {\em
quasi-stationary\/} point of $\overline{s}$ at the edge of the
MSR, which we call the MSR {\em apex}. The apex is characterized
by the shortest distance between the sets of minimums of
$\overline{s}$ in the $\overline{d}$- and $\sigma$-directions. The
two sets do not intersect but are extremely close to each other at
the apex. The surface of the average stretch function values in
the apex neighborhood consists solely of elliptic
points,\footnote{Any point on a regular surface is always of one
of the following four types: planar (example: any point on a
plane), elliptic (examples: any point on an ellipsoid, peaks and
pits), parabolic (examples: any point on a cylinder, ridges and
channels), and hyperbolic (examples: any point on a hyperboloid,
passes). A function of two arguments has an extremum at its
stationary point if the corresponding point on a surface of its
values is elliptic.} but the minimal deformation of the surface
towards the potential intersection appears to result in a unique
parabolic point.

The points corresponding to distance distributions of all random
graphs with power-law node degree distributions lie in the MSR. In
addition, the Internet distance distribution is located in a very
{\em close neighborhood\/} to the MSR apex. Even a stronger
statement is valid: $\gamma=2.1$ is the value of $\gamma$
corresponding to the distance distribution that is {\em closest\/}
to the apex, compared to all other values of $\gamma$. The ${\cal
G}_{n,p}$ graphs are far away both from the MSR and from its apex.

The phenomena outlined above appear to be a reflection of
existence of a certain link between the Internet topology and the
analytical structure of the average TZ stretch function. This is
quite unexpected since the Internet, as we know it today, seems to
have nothing to do with stretch, in general, and with the TZ
stretch, in particular. That is why these effects cannot be fully
interpreted within the set of ideas we operate with in this paper.
Although, see Section \ref{sec:conclusions} for some hints towards
possible explanations.

In a recent work dedicated to a specific subject, \cite{GaNe03},
Gavoille and Neh\'{e}z raise a very important general issue of
application of results in ``theoretical'' routing to {\em
realistic\/} networks.\footnote{They also question what realistic
networks are. We believe that this question is being actively
answered in the work discussed in Section
\ref{sec:scale-free_networks}.} To the best of our knowledge, our
work is among the first ones trying to create a link between
routing and realistic scale-free networks. The principal result of
this paper showing that the TZ stretch on Internet-like graphs is
low, opens a well-defined path for the future work in this area,
as further discussed in Sections \ref{sec:conclusions} and
\ref{sec:future_work}.

\section{Stretch}

Both the Cowen and TZ schemes are very simple. They involve four
separate components: the landmark set (LS) construction procedure,
routing table construction, labelling, and routing itself. The TZ
scheme differs from the Cowen scheme by improving just the first
part; the other three are the same. We remind the outline of the
TZ scheme below.

The scheme operates on any undirected graph $G=(V,E)$ with
positive edge weights. Let $n = |V|$ be the graph size,
$\delta(u,v)$ be the distance between a pair of nodes $u,v \in V$,
$L$ be the LS, $L(v)$ be a landmark node closest to node $v \in
V$, and $C(v)$ be $v$'s cluster defined for $\forall v \in V$ as a
set of all nodes $c$ that are closer to $v$ than to their closest
landmarks,
\begin{equation}\label{cluster_def}
    C(v) = \big\{ \; c \in V \; \big| \; \delta(c,v) < \delta(c,L(c))
    \;\big\}.
\end{equation}
Clusters are similar to the Voronoi diagrams but they can
intersect. If $l \in L$, then $L(l) = l$ and $C(l) = \emptyset$ by
definition. If $L$ is empty, then for $\forall v \in V$, $L(v) =
\emptyset$ and $C(v) = V$.

The TZ LS construction algorithm interactively selects landmarks
from the set of large-cluster nodes $W$. At the first iteration,
$W=V$ and every node $w \in W$ is selected to be a landmark with a
specific uniform probability $q/n$ with $q = (n/\log n)^{1/2}$.
The expected LS size after the first iteration is $q$. At the
subsequent iterations, $W$ is redefined to be a set of nodes that
have clusters of size greater than a specific threshold $\tilde{q}
= 4n/q$,
\begin{equation}\label{large_cluster_def}
    W = \big\{ \; w \in V \; \big| \; |C(w)| > \tilde{q} \;\big\} ,
\end{equation}
and additional portions of landmarks are selected from $W$ with a
uniform probability $q/|W|$. The iterations proceed until $W$ is
empty.

Every node $v \in V$ calculates then its outgoing port for the
shortest path to every $l \in L$ and every $c \in C(v)$. This is
the routing information that is stored locally at $v$. As one can
see, the essence of the LS construction procedure is the right
balance between the LS and cluster sizes (or, effectively, between
$q$ and $\tilde{q}$). The cluster sizes are upper-bounded by
definition (\ref{large_cluster_def}), and the involved part of the
proof is to demonstrate that the algorithm terminates with a
proper limit for the expected LS size, which turns out to be $2 q
\log n$. This guarantees the overall local memory upper bound of
$O(n^{1/2}\log^{1/2}n$).

The label of node $v$ (used as its destination address in packet
headers) is then a triple of its ID, the ID of its closest
landmark $L(v)$, and the local ID of the port at $L(v)$ on the
shortest path from $L(v)$ to $v$. With these labels, routing of a
packet destined to $v$ at some (intermediate) node $u$ occurs as
follows: if $v = u$, done; if $v \in L \cup C(u)$, the outgoing
port can be found in the local routing table at $u$; if $u =
L(v)$, the outgoing port is in the destination label in the
packet; otherwise, the outgoing port for the packet is the
outgoing port to $L(v)$---the $L(v)$ ID is in the label and the
outgoing port for it can be found in the local routing table. The
demonstrations of correctness of the algorithm and that the
maximum stretch is 3 are straightforward
(\cite{cowen01,ThoZwi01b}).

\subsection{Analytical results}\label{sec:analytical_results}

In this section, we provide analytical expressions for the TZ
stretch distribution on a small-world graph with a given distance
distribution, in general, and with the Gaussian distance
distribution, in particular.

We start with the following assumption drastically simplifying the
analysis:
\begin{assumption}\label{ass:one_iteration}
Only the first iteration of the LS construction algorithm is
considered.
\end{assumption}
There are two justifications making this assumption reasonable.
First, as shown both in~\cite{ThoZwi01a} and below in
Claim~\ref{claim:average-cluster}, the first iteration guarantees
that the {\em average\/} cluster size is below~$n/q$; the
subsequent iterations guarantee that {\em all\/} cluster sizes are
upper-bounded by $4n/q$. Therefore, the error introduced by this
assumption for the {\em average\/} stretch is small as we see in
the next section. The second justification making the error
particularly small is that we consider small-world graphs which
have very short average distances and narrow distance
distributions. Indeed, if there are no long distances in a graph,
then even after just the first iteration, the majority of clusters
are small. The error introduced by the assumption is related to
the difference between the expected LS size $q$ after the first
iteration and the average LS size observed in simulations, which
is reported in the next section.

The fact that ratio between the expected LS size and the graph
size is infinitesimally small for large graphs,
$q/n\xrightarrow[n\to\infty]{}0$, makes the following claim true:
\begin{claim}\label{claim:no_difference}
The difference between the distance distribution in $G$ and
distance distribution in its subgraphs $\tilde{G}$ induced by $V
\setminus \tilde{L}$, $\forall \tilde{L} \subset L$, can be
neglected for large graphs.
\end{claim}

For the rest of this section, we let $q$ denote the actual size of
the LS, $q = |L|$. We also denote the distance p.d.f.\ and c.d.f.\
by $f(d)$ and $F(d)$ respectively. With $D$ being the graph
diameter, we allow $d = 0 \ldots D$, where $f(0) = 1/n$ is the
probability of zero-distance from a random node to itself. In some
places below, we also refer to the continuous limit approximation
(that is, to the assumption that~$f(d)$ is continuous), but we
explicitly avoid using it in the evaluations of the next
section.\footnote{Note, however, that the discrete case can often
be closely approximated by the continuous case. Indeed, recall
that as soon as~$f(d)$ is sufficiently smooth and a sufficient
number of its first derivatives are small enough at the interval
boundaries $[0,D]$, which is the case for the Gaussian form
of~$f(d)$ we eventually select, then, according to the
Euler-Maclaurin sum formula, the sum becomes indistinguishable
from the integral over the same interval,
$\sum_{d=0}^Df(d)\to\int_0^Df(d)\,\dif d$.} With the above
notations and Assumption~\ref{ass:one_iteration}, we are ready to
formulate the following claim:
\begin{claim}\label{claim:gi}
The p.d.f.\ for the distance between a random node and its $i$'th
closest landmark is given by
\begin{equation}\label{gi}
    g_i(d) = c_i F(d)^{i-1} f(d)
    (1-F(d))^{q-i}
\end{equation}
with normalization coefficients $c_i=i\binom{q}{i}$ in the
continuous limit.
\end{claim}
Expression (\ref{gi}) is intuitively expected since $F(d)^{i-1}$
approximates the probability that $i-1$ landmark nodes are closer
than $d$, and $(1-F(d))^{q-i}$ approximates the probability that
the rest of landmark nodes are farther than $d$ (cf.\ the {\em
order statistics},~\cite{Ross00}). For a more rigorous proof of
Claims \ref{claim:no_difference} and \ref{claim:gi} above, see
Appendix \ref{sec:proofs:gi}.

Given the p.d.f.\ for the distance to the closest landmark
$g_1(d)$ in (\ref{gi}), one can easily prove (see Appendix
\ref{sec:proofs:average-cluster}) the following claim:
\begin{claim}\label{claim:average-cluster}
The average cluster size $\overline{|C|} \leq
n/(q+1)$.\footnote{Note that $q+1$ instead of $q$ in the
denominator guarantees the right bound for the case of the empty
LS, $q=0$.}
\end{claim}

We next denote by $g(d)$ the p.d.f.\ for the average distance to
{\em all\/} landmarks,
\begin{equation}\label{g-def}
    g(d) = \frac{1}{q} \sum_{i=1}^q g_i(d).
\end{equation}
Since landmarks are just some $q$ random nodes, $g(d)$ is
equivalent to $f(d)$. In the continuous limit,\footnote{In the
discrete case and with~$f(d)$ being Gaussian, $g(d)$ is still
virtually identical to~$f(d)$ because of the Euler-Maclaurin sum
formula.} we have
\begin{equation}
    g(d) = f(d)\frac{1}{q}
    \sum_{i=1}^q i \binom{q}{i} F(d)^{i-1} (1-F(d))^{q-i} = f(d).
\end{equation}

Letting $w$ be the source node and $v$ be the destination, we fix
the notation for the following three random variables:
\begin{align}
  x &= \delta(w,L(v)), & \mathrm{p.d.f.}&=g(x), \label{x-def}\\
  y &= \delta(v,L(v)), & \mathrm{p.d.f.}&=g_1(y), \label{y-def}\\
  z &= \delta(w,v),    & \mathrm{p.d.f.}&=f(z). \label{z-def}
\end{align}
With these notations, the random variable for the approximate
stretch is
\begin{equation}\label{stretchval-approx}
    s^\ast(x,y,z) = \frac{x+y}{z}.
\end{equation}
This expression for stretch is approximate for two reasons. First,
it does not account for stretch-1 paths to destinations in the
local cluster. Second, it does not incorporate the {\em shortcut
effect}. Recall that the Cowen routing algorithm is such that if
destination $v \not\in L$ and if a message on its way to $L(v)$
passes some node $u\;\big|\; v \in C(u)$, then the message never
reaches $L(v)$ but goes along the shortest path from $u$ to $v$.
In Appendix \ref{sec:proofs:shortcut}, we justify the following
claim:
\begin{claim}\label{claim:shortcut}
The stretch-1 and shortcut paths can be approximated by the
following correction to $s^\ast$ in (\ref{stretchval-approx}):
\begin{equation}\label{stretchval-shortcut}
    s(x,y,z) =
    \begin{cases}
        1 & \text{if \; $z<y$,}\\
        1 & \text{if \; $z<x$,}\\
        \frac{x+y}{z} & \text{otherwise.}
    \end{cases}
\end{equation}
\end{claim}

Our problem now is to find the joint p.d.f.~$t(x,y,z)$ for
$s(x,y,z)$. If $x$, $y$, and $z$ were independent random
variables, then $t(x,y,z)$ would be given by $g(x)g_1(y)f(z)$.
They are not independent by
definitions~(\ref{x-def})-(\ref{z-def}), which result in the
triangle inequality,
\begin{equation}\label{triangle-ineq}
    |x-y| \leq z \leq x+y.
\end{equation}
Furthermore, there can be some other correlations in the distance
matrix. To proceed, we make the following assumption:
\begin{assumption}\label{ass:no-corr}
There are no correlations in the distance matrix, other than those
associated with the triangle inequality.
\end{assumption}
With this assumption, we can prove (see
Appendix~\ref{sec:proofs:triangle}) the following claim:
\begin{claim}\label{claim:triangle}
The stretch p.d.f.\ is given by
\begin{equation}\label{triangle}
    t(x,y,z) = \frac{g(x)g_1(y)f_t(x,y,z)}{F(x+y)-F(|x-y|)}, \quad
    f_t(x,y,z) =
    \begin{cases}
        f(z) & \text{if \; $|x-y| \leq z \leq x+y$,}\\
        0    & \text{otherwise.}
    \end{cases}
\end{equation}
\end{claim}
This claim is intuitively expected---the triangle inequality
(\ref{triangle-ineq}) just cuts a corresponding portion out from
$f(z)$ with the proper normalization coefficient.

The average stretch and the stretch distribution are now
\begin{eqnarray}
  \overline{s} &=& \sum_{x,y,z=0}^D
  s(x,y,z) t(x,y,z), \label{stretch-average} \\
  \rho(\varsigma) &=& \sum_{\substack{x,y,z=0 \\
  s(x,y,z) = \varsigma}}^D t(x,y,z).
  \label{stretch-distrib}
\end{eqnarray}
In the above expression for the stretch distribution
$\rho(\varsigma)$, the summation is over such values of $x$, $y$,
and $z$ that their transformation according to
(\ref{stretchval-shortcut}) yields~$\varsigma$.

Equations~(\ref{stretch-average}) and~(\ref{stretch-distrib}) are
out final analytical results that we require for the numerical
evaluations of the next section. Of particular note is that the
stretch distribution and average depend only on~$f(d)$ and~$q$.

At this point, however, we may try to substitute any specific form
of the distance distribution into~(\ref{stretch-average})
and~(\ref{stretch-distrib}). As discussed in the introduction, we
are interested in the Gaussian distance distribution with the
average distance~$\overline{d}$ and standard deviation
(width)~$\sigma$,
\begin{equation}\label{gaussian-pdf}
    f(d) = \frac{1}{\sigma\sqrt{2\pi}} \; e^{\textstyle -\frac{1}{2}
    \left( \frac{d-\overline{d}}{\sigma} \right)^2}.
\end{equation}
Assuming that the distribution is continuous,\footnote{In fact,
the Gaussian distribution {\em is\/} continuous by definition.
According to the de Moivre-Laplace theorem, it is an asymptotic
form of the binomial distribution,
$\binom{D}{d}\vartheta^d(1-\vartheta)^{D-d}
\xrightarrow[D\to\infty]{}
(\sigma\sqrt{2\pi})^{-1}e^{-(d-\overline{d})^2/(2\sigma^2)}$ with
$\vartheta = \overline{d}/D$ and $\sigma^2 =
D\vartheta(1-\vartheta)$. Although the values of $D \sim 13$,
$\overline{d} \sim 3.6$, and $\sigma \sim 0.9$ observed in the
Internet make this approximation essentially invalid for
analytical purposes, we can still use (\ref{gaussian-pdf}) with
discrete $d$ for {\em numerical\/} computations.} we can express
the distance c.d.f.\ via the error function,
\begin{equation}\label{gaussian-cdf}
    F(d) = \frac{1}{2}\left[1+\erf\left(\frac{d-\overline{d}}{\sigma\sqrt{2}}\right)\right].
\end{equation}
The average stretch becomes the following integral:
\begin{equation}
    \overline{s} = \iiint s(x,y,z) t(x,y,z) \,
    \dif x \dif y \dif z,
\end{equation}
which, after a series of substitutions, transforms to
\begin{equation}\label{stretch-superlong}
    \overline{s}(\overline{d},\sigma) =
    \frac{2^{2-q}q}{\sigma^3(2\pi)^{3/2}}
    \iint
    \frac{e^{-\frac{1}{2\sigma^2}
            \left[(x-\overline{d})^2 + (y-\overline{d})^2\right]}
        \left[1-\erf\left(\frac{y-\overline{d}}{\sigma\sqrt{2}}\right)\right]^{q-1}}
    {\erf\left(\frac{x+y-\overline{d}}{\sigma\sqrt{2}}\right)
        - \erf\left(\frac{|x-y|-\overline{d}}{\sigma\sqrt{2}}\right)}
    \, \dif x \dif y
    \int\limits_{|x-y|}^{x+y}
    e^{\textstyle -\frac{1}{2}\left(\frac{z-\overline{d}}{\sigma}\right)^2}
    s(x,y,z) \, \dif z.
\end{equation}
Unfortunately, we cannot evaluate even the inner-most integral in
any special functions known either to us or to Gradstein-Ryzhik
\cite{gradstein-ryzhik}. Therefore, we retreat to numerical
evaluations of~(\ref{stretch-average}) and~(\ref{stretch-distrib})
in the explicitly discrete case.

\subsection{Numerical results and simulations}\label{sec:numerical_results_and_simulations}

In our numerical evaluations of (\ref{stretch-average}) and
(\ref{stretch-distrib}), $x$, $y$, and $z$ (defined in
(\ref{x-def})-(\ref{z-def})) are integer variables with the
following ranges:
\begin{eqnarray}
  x,y &=& 1 \ldots D, \label{xy-range}\\
  z   &=& \max\left(1,|x-y|\right) \ldots \min\left(D, x+y\right), \\
  D   &=& \left[\,\overline{d}\,\right] +
  \left\lceil 10\sigma\sqrt{2} \right\rceil,
\end{eqnarray}
where
$\left[\,\overline{d}\,\right]\equiv\mathrm{round}\left(\overline{d}\right)$
and diameter $D$ becomes a distance distribution cutoff parameter,
$f(d) \ll 1$, $\forall d>D$ since $f(D)/f(\overline{d})\sim
e^{-100}$. We do not have any singularities that we have to deal
with and that are present, for example,
in~(\ref{stretch-superlong}). The TZ LS size~$q$ is rounded:
\begin{equation}
    q=\left[\,\sqrt{\frac{n}{\log_2n}}\,\right],
\end{equation}
and all distance distributions are explicitly normalized,
e.g.~$f(d)$ from (\ref{gaussian-pdf}) is taken to be
\begin{equation}\label{gaussian-pdf-norm}
    f(d)= c\,e^{\textstyle -\frac{1}{2}
    \left( \frac{d-\overline{d}}{\sigma} \right)^2},\quad
    \text{$c$ is such that} \sum_{d=1}^{D} f(d) = 1,
\end{equation}
and distributions~$g(x)$ and~$g_1(y)$ are explicitly normalized as
well.

For the simulation part, we use our TZ scheme simulator on the
graphs produced by the PLRG generator. For a given parameter set,
all the data is averaged over $10$ random graphs. All average
graph sizes $n$ are between $10,000$ and $11,000$ unless mentioned
otherwise.

\subsubsection{Distance distribution}

We have to stress here that the stretch distribution is a function
of the distance distribution and the graph size only. Therefore,
all we have to verify for our results having practical value is
that both the distance distribution we use for the analysis and
the distance distribution in the generated graphs are close to the
distance distribution observed in the Internet.

\begin{figure*}[t]
\centerline{\epsfig{file=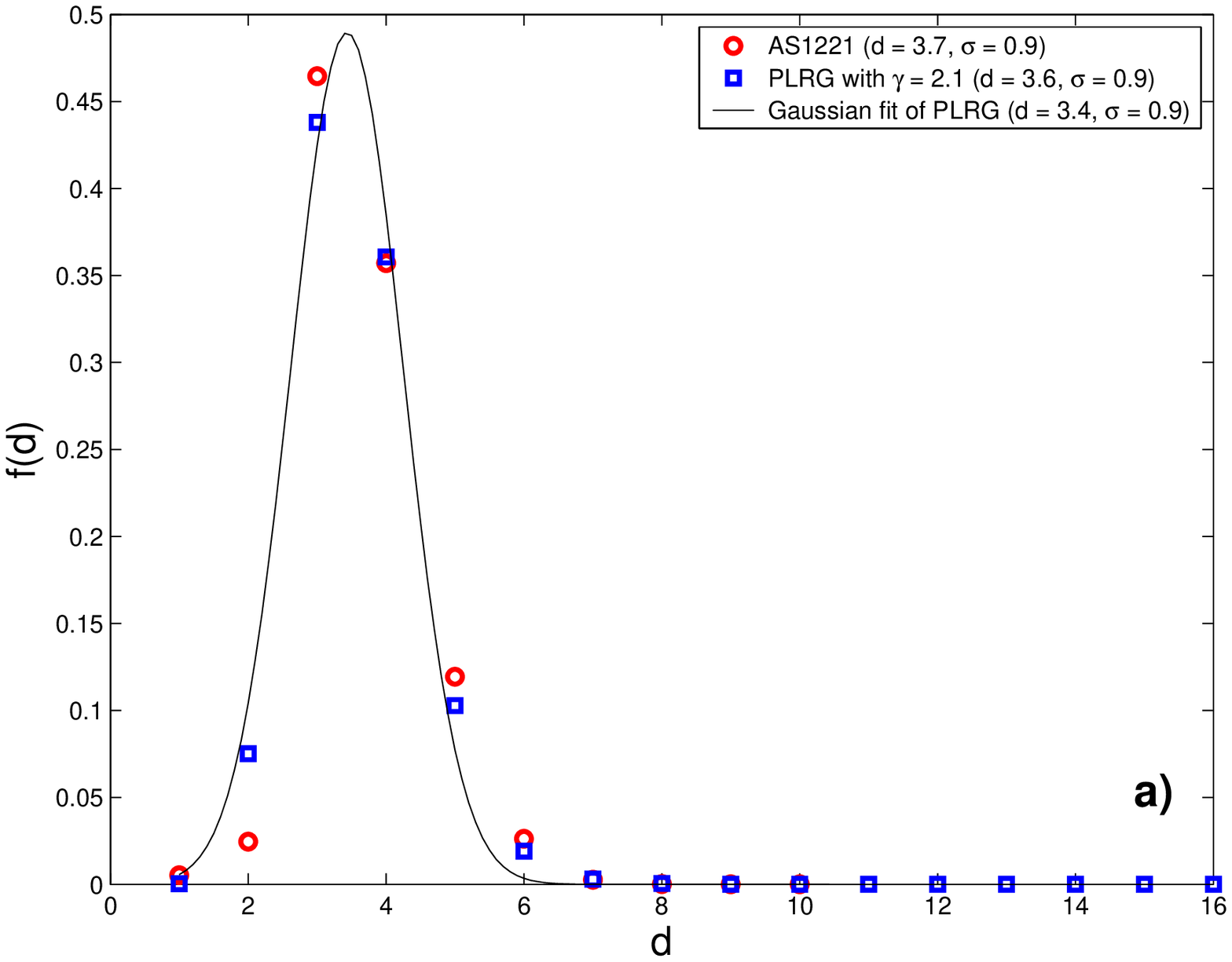,width=3in} \hspace*{0.5in}
  \epsfig{file=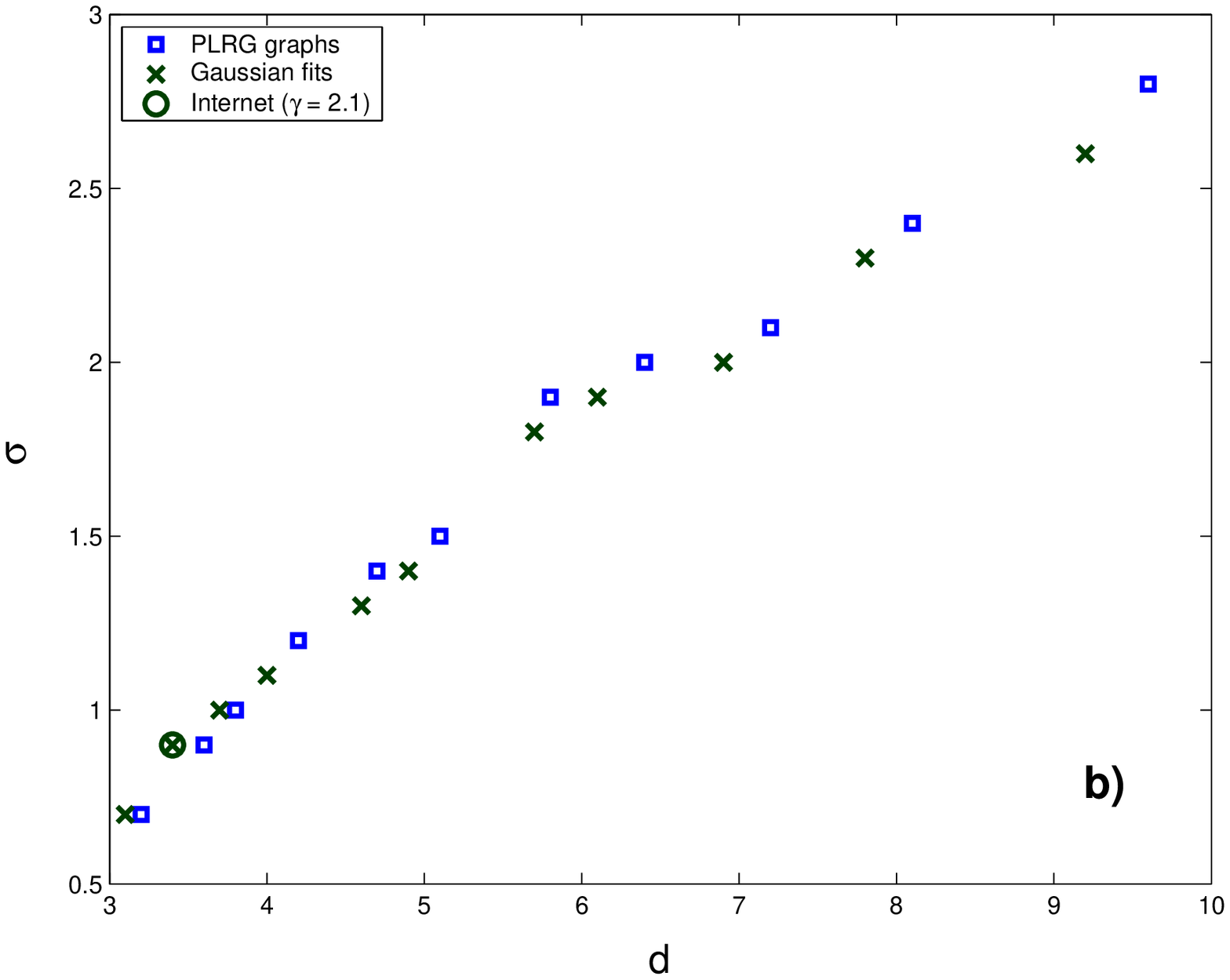,width=3in}}

\hspace*{0.2in}

\centerline{\epsfig{file=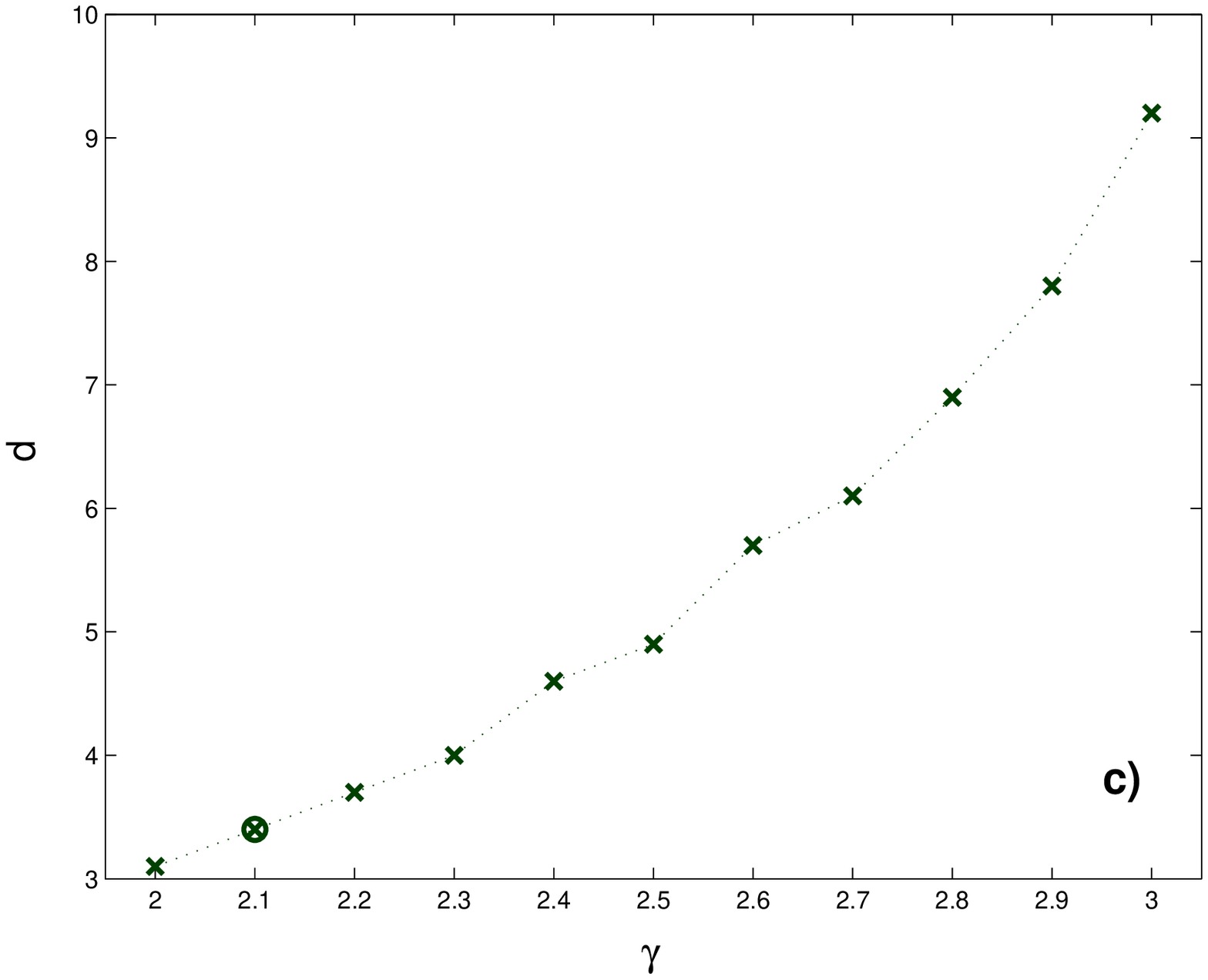,width=3in} \hspace*{0.5in}
  \epsfig{file=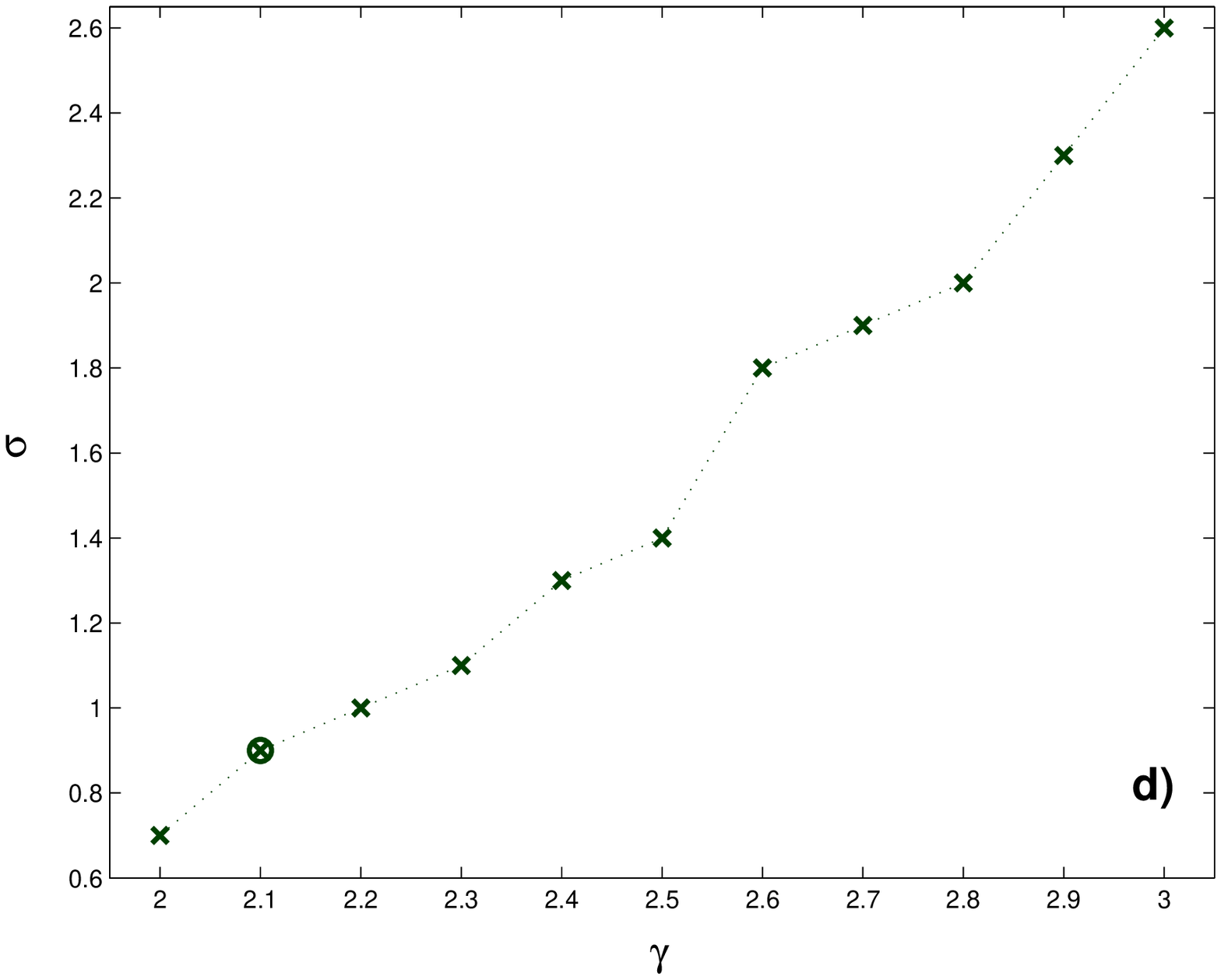,width=3in}}

\caption{\textbf{a)} The distance distributions. The red circles
represent the distance distribution from a typical AS (AS~\#1221)
averaged over a period of approximately March-May, 2003. (The
source of data is \cite{huston-bgp-site}; for other measurements,
see \cite{VaPaVe02,BroNeCla02}.)  The mean and the standard
deviation is $3.7$ and $0.9$ respectively. The distance
distribution in PLRG-generated graphs with $\gamma=2.1$ is shown
by blue squares. The standard deviation is the same as before, the
mean is $3.6$. The solid line is the Gaussian fit of the PLRG
distribution, $\overline{d}=3.4$ and $\sigma=0.9$. \textbf{b)} The
means and standard deviations (blue squares) of distance
distributions in PLRG-generated graphs with
$\gamma=2.0,2.1,\ldots,3.0$ (from left to right), and the
corresponding values of $\overline{d}$ and $\sigma$ (green
crosses) in their Gaussian fits. The fitted values of
$\overline{d}$ and $\sigma$ as functions of $\gamma$ are shown in
\textbf{(c)} and \textbf{(d)} respectively. The Internet value of
$\gamma=2.1$ is circled in (b)-(d).}

\label{fig:dd}
\end{figure*}

Based on the experiments performed in \cite{TaGoJaShWi02}, one can
expect that the distance distribution in PLRG-generated graphs
should be close to the one in the Internet. We find that it is
indeed so. See Fig.\ \ref{fig:dd}(a) for details.

Then we proceed as follows. Paying a special attention to the
value of the node degree distribution exponent $\gamma$ equal to
$2.1$, which is observed in the Internet, we generate series of
graphs with $\gamma$ ranging from $2$ to $3$, and calculate their
distance distributions. We fit these distributions by explicitly
normalized Gaussians (\ref{gaussian-pdf-norm}) yielding values of
$\overline{d}$ and $\sigma$ that we use in numerical evaluations
of our analytical results. For fitting, we use the standard
non-linear least squares method. All fits are very good: the
maximum SSE we observe in our fits is $0.003$ and the minimum
R-square is $0.9905$.

The values of $\overline{d}$ and $\sigma$ in fitted Gaussians are
slightly off from the means and standard deviations of distance
distributions in generated graphs as depicted in Fig.\
\ref{fig:dd}(b). In fact, Fig.\ \ref{fig:dd}(b) is a parametric
plot of $\sigma(\overline{d})$ with $\gamma$ being a parameter. We
observe almost linear relation between $\overline{d}$ and $\sigma$
with such parametrization. Note that almost linear relation
between the distance c.d.f.\ center and width parameterized by
$\gamma$ is analytically obtained in \cite{DoMeSa03a}. We further
discuss this subject in Section \ref{sec:minimum}. In Fig.\
\ref{fig:dd}(c,d), we show fitted $\overline{d}$ and $\sigma$ as
functions of $\gamma$ (cf.\ the results in
\cite{DoMeSa03a,ChLu03}).

Average graph sizes for different values of $\gamma$ are slightly
different but dependence of $\overline{d}$ and $\sigma$ on $n$
(not shown) is negligible compared to their dependence on
$\gamma$. This is in agreement with \cite{DoMeSa03a,ChLu03}.

\subsubsection{Stretch distribution}\label{sec:stretch_distribution}

\begin{figure*}[t]
\centerline{\epsfig{file=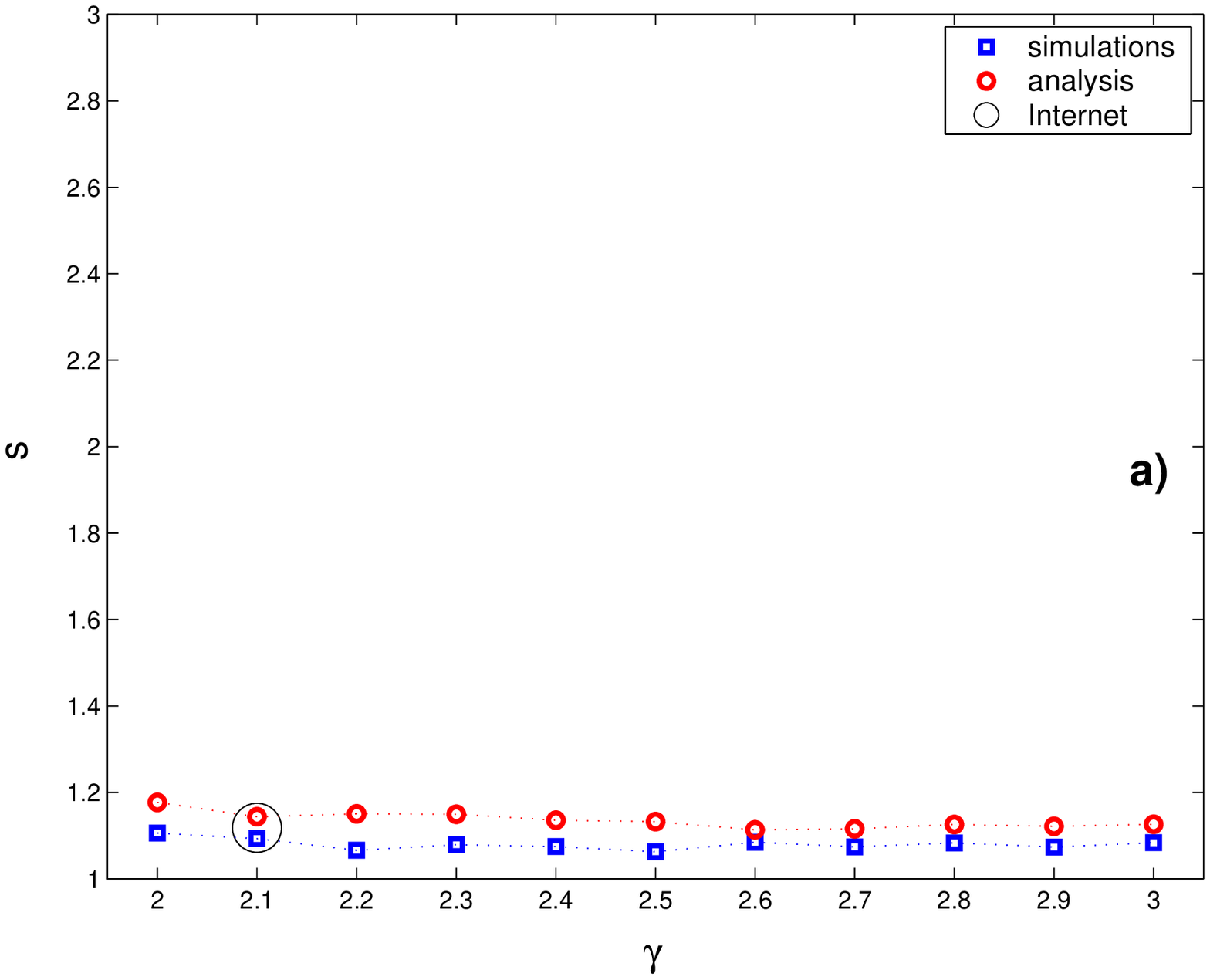,width=3in} \hspace*{0.5in}
  \epsfig{file=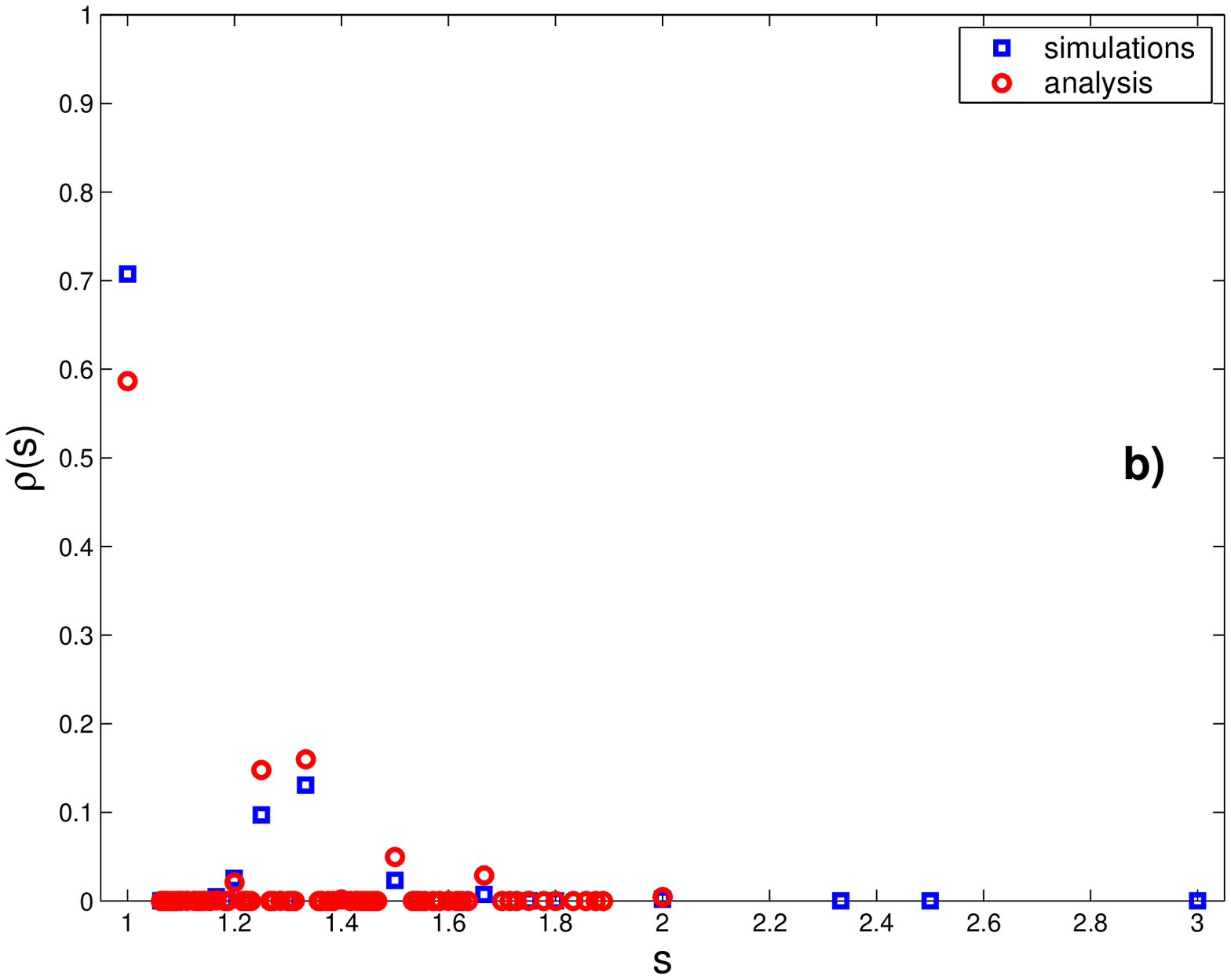,width=3in}}

\hspace*{0.2in}

\centerline{\epsfig{file=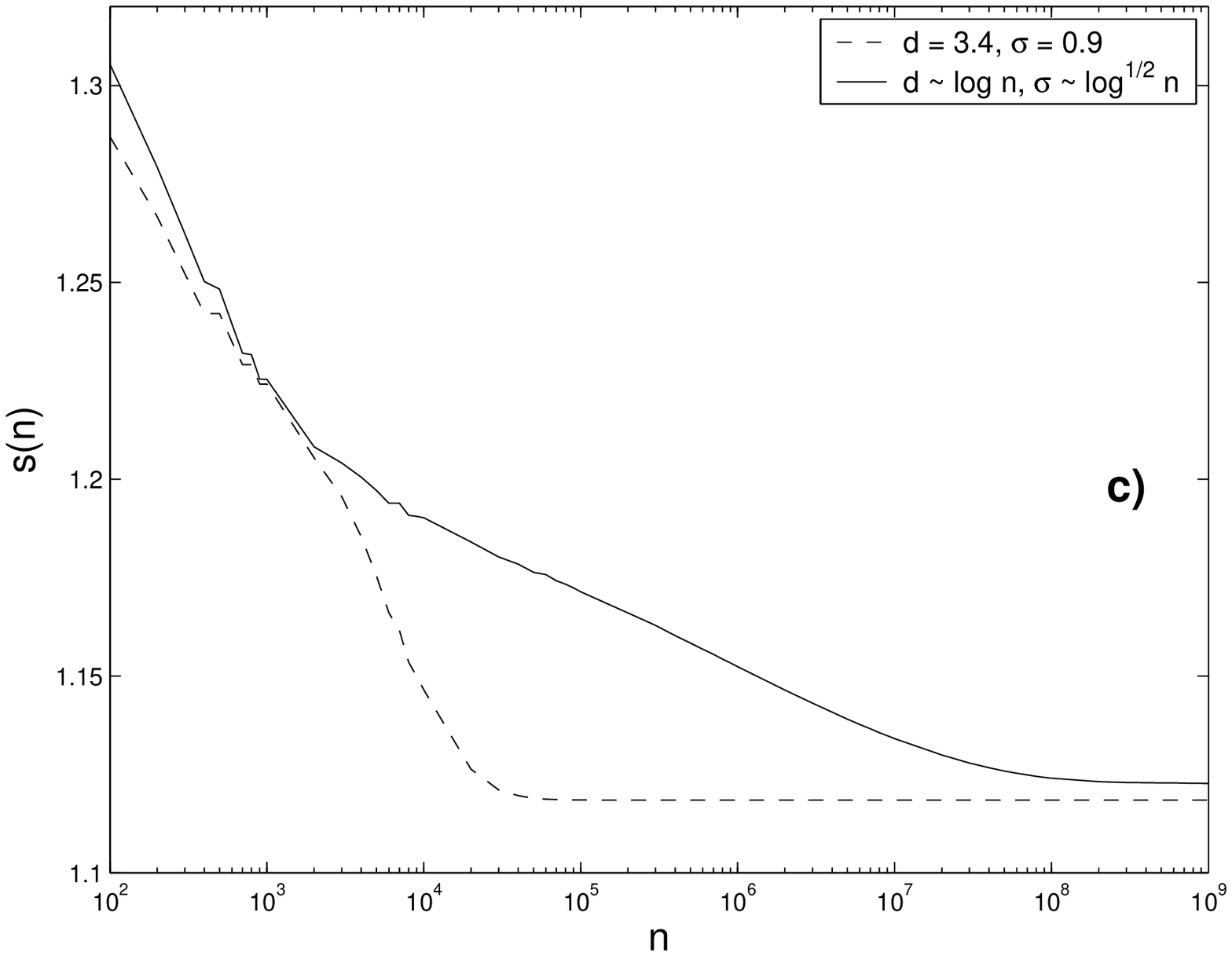,width=3in} \hspace*{0.5in}
\epsfig{file=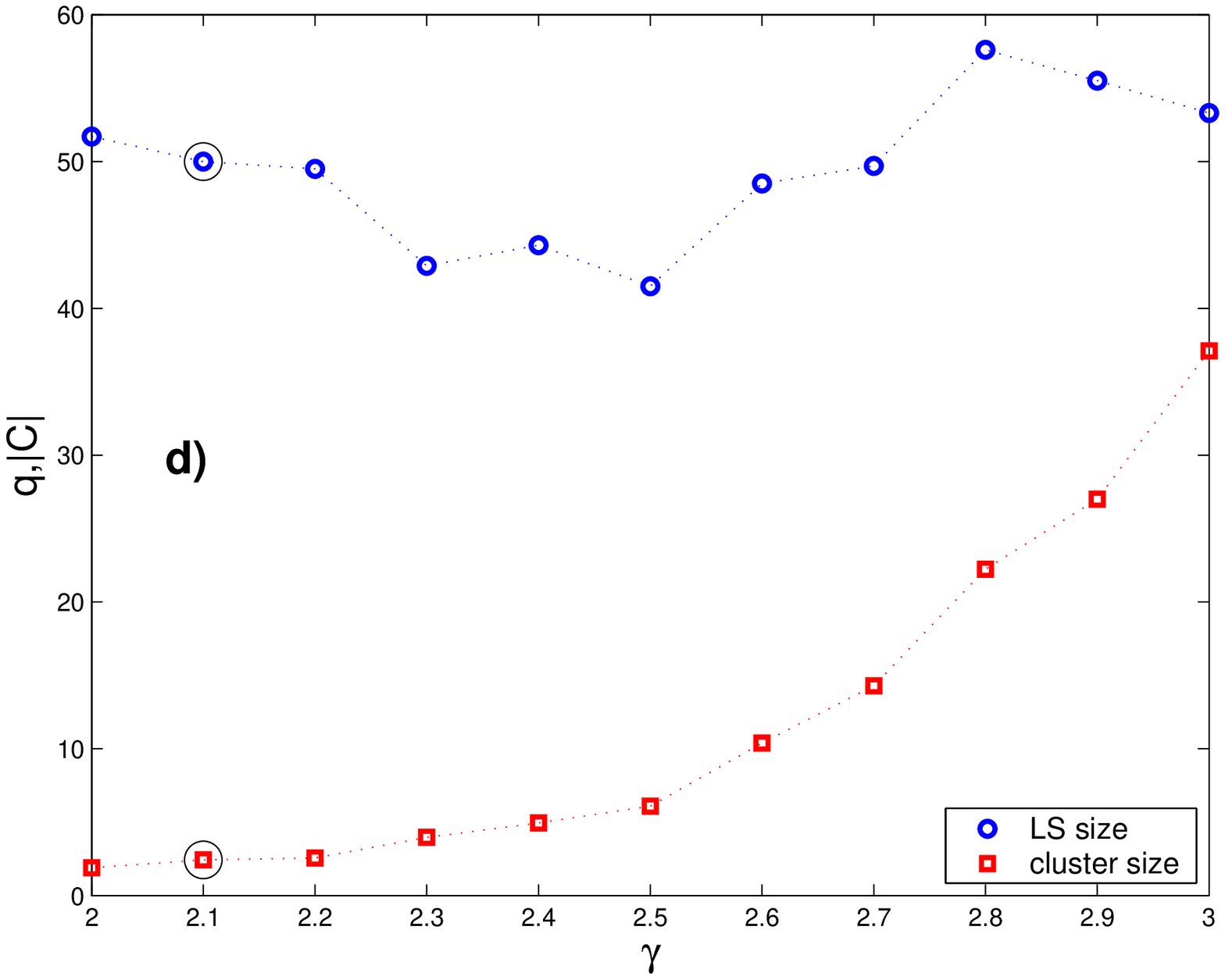,width=3in}}

\caption{\textbf{a)} The analytical results (red circles) and
simulation data (blue squares) for the average TZ stretch as a
function of $\gamma$. \textbf{b)} The analytical results (red
circles) and simulation data (blue squares) for the TZ stretch
distribution with $\gamma=2.1$. \textbf{c)} The analytical data
for the average stretch as a function of the graph size. The
dashed line corresponds to the case when the distance distribution
parameters $\overline{d}$ and $\sigma$ are fixed to the values
observed in the Internet. The solid line presents the data when
$\overline{d}$ and $\sigma$ scale according to the DGM model.
\textbf{d)} The simulation data for the LS (blue circles) and
cluster (red squares) sizes. In the Internet case, $\gamma=2.1$,
the average graph size in simulations is 10,687, the average LS
size is 50.0, and the average cluster size is 2.43.}

\label{fig:sd}
\end{figure*}

\begin{table}[t]
  \centering
  \caption{The top ten stretch values and
  percentage of paths associated with them.}
  \label{table:stretch_top10}
  \begin{tabular}{|c|c|c|}
    \hline
    Stretch & Analysis (\%) & Simulations (\%) \\
    \hline\hline
    $1$   & $58.7$ & $70.8$  \\ \hline
    $4/3$ & $16.0$ & $13.1$  \\ \hline
    $5/4$ & $14.8$ & $9.71$  \\ \hline
    $3/2$ & $4.95$ & $2.33$  \\ \hline
    $5/3$ & $2.88$ & $0.731$ \\ \hline
    $6/5$ & $2.10$ & $2.54$ \\ \hline
    $2$   & $0.434$ & $0.210$ \\ \hline
    $7/5$ & $0.173$ & $6.77 \times 10^{-2}$ \\ \hline
    $7/6$ & $5.20 \times 10^{-2}$ & $0.460$ \\ \hline
    $8/7$ & $3.01 \times 10^{-4}$ & $7.42 \times 10^{-2}$ \\ \hline
  \end{tabular}
\end{table}

We obtain a very close match between the simulations and analysis
of the average TZ stretch and stretch distribution. The average
stretch as a function of $\gamma$ is shown in
Fig.~\ref{fig:sd}(a). For the Internet-like graphs, $\gamma=2.1$,
the average stretch we observe in simulations is $1.09$ and the
average stretch given by (\ref{stretch-average}) with $f(d)$ in
(\ref{gaussian-pdf-norm}), with $\overline{d}=3.4$ and
$\sigma=0.9$, is $1.14$.\footnote{The sources of the small error
are in Assumptions \ref{ass:one_iteration}, \ref{ass:no-corr}, and
in approximations of Claim \ref{claim:shortcut}.} Thus, we find
that the average stretch is {\em very low}.

Furthermore, while both the average distance and distance
distribution width in power-law graphs do depend on $\gamma$ (cf.\
Fig.~\ref{fig:dd}(c,d)), the average stretch {\em does not}. We
delay the discussion of this topic until Section
\ref{sec:minimum}.

The stretch distributions obtained both analytically,
(\ref{stretch-distrib}), and in simulations are shown in
Fig.~\ref{fig:sd}(b). The sets of significant stretch values (that
is, stretch values having noticeable probabilities) match between
the analysis and simulations. The top ten stretch values
corresponding to virtually 100\% of paths are presented in
Table~\ref{table:stretch_top10}.

We notice that a majority of paths (up to $\sim 71\%$ according to
the simulations) are {\em shortest}. There are only a very few
significant stretch values for the rest of paths. All the
significant stretch values are below $2$.

The small amount of stretch values with noticeable probabilities
is due to the narrow width of the distance distribution. Indeed,
in $\sim 86\%$ cases, two random nodes are either 3 or 4 hops away
from each other. That is, the probability for $x$ or $z$ to be
either $3$ or $4$ is $\sim0.86$, see Fig.~\ref{fig:dd}(a). In
$\sim 82\%$ cases, a random node is just one hop away from its
closest landmark, $g_1(1)\sim0.82$. This explains why
stretch-$4/3$ ($x=3$, $y=1$, and $z=3$) and stretch-$5/4$ ($x=4$,
$y=1$, and $z=4$) paths are most probable among stretch $s>1$
paths in Table~\ref{table:stretch_top10}.

In Fig.~\ref{fig:sd}(c), the analytical results for the average
stretch as a function of the graph size are shown. Note that
dependence on $n$ in (\ref{stretch-average}) is only via the LS
size $q$. We present data for the case when $\overline{d}$ and
$\sigma$ are fixed at their values observed in the Internet, and
the case when they are allowed to scale as in the DGM model. In
both cases, the average stretch slowly {\em decreases\/} as the
network grows, although this decrease is spread over multiple
orders of magnitude of $n$ and the stretch change is confined to a
narrow region between $1.3$ and $1.1$. We also notice that after a
certain point, the stretch stops decreasing. Although it becomes
very small, it does not reach its minimal value 1.

Finally, in Fig.~\ref{fig:sd}(d), we report the simulation data on
the average cluster and LS sizes. We notice that they are well
below their bounds. The average cluster size growth similar to the
growth of the average distance, cf.\ Fig.~\ref{fig:dd}(c), is
expected.

Recall that the sum of the cluster and LS sizes in the TZ scheme
is the number of records in the local routing table. We see that
for the Internet-like graphs, $n\sim10^4$, $\gamma\sim2.1$, this
sum is $\sim52$.

\subsubsection{${\cal G}_{n,p}$ graphs}\label{sec:gnp}

Looking at Figs.~\ref{fig:sd}(a,c), one may be tempted to assume
that the average stretch just moderately depends on $n$ and does
not depend on either $\overline{d}$ or $\sigma$ for a wide class
of random graphs.

To demonstrate that this is incorrect, we consider the most common
class of random graphs, ${\cal G}_{n,p}$. We take $n \sim 10^4$
and choose $p$ to match approximately the Internet average
distance ($p \sim 1.3\times10^{-3}$) and average node degree ($p
\sim 5.7\times10^{-4}$). The analytical and simulation results for
the average stretch in these two cases are presented in
Table~\ref{table:gnp}. We find that the average stretch is
substantially higher than in the case of random graphs with
power-law node degree distributions.

\begin{table}
  \centering
  \caption{The average TZ stretch on the ${\cal G}_{n,p}$ graphs.}
  \label{table:gnp}
  \begin{tabular}{|c|c|c|c|c|c|c|}
    \hline
    $n$ & $p$ & Avg. degree $\overline{k}$ & $(\overline{d},\sigma)$ in graphs &
    $(\overline{d},\sigma)$ in Gaussian fits & $\overline{s}$ (analysis) &
    $\overline{s}$ (simulations) \\
    \hline\hline
    $10^4$ & $1.3\times10^{-3}$ & 13 & $(3.9,0.6)$ & $(3.9,0.5)$ &
    $1.51$ & $1.60$ \\
    \hline
    $10^4$ & $5.7\times10^{-4}$ & $5.7$ & $(5.5,0.9)$ &
    $(5.6,0.8)$ & $1.37$ & $1.50$ \\
    \hline
  \end{tabular}
\end{table}

\section{The apex}\label{sec:minimum}

\begin{figure*}[t]
\centerline{\epsfig{file=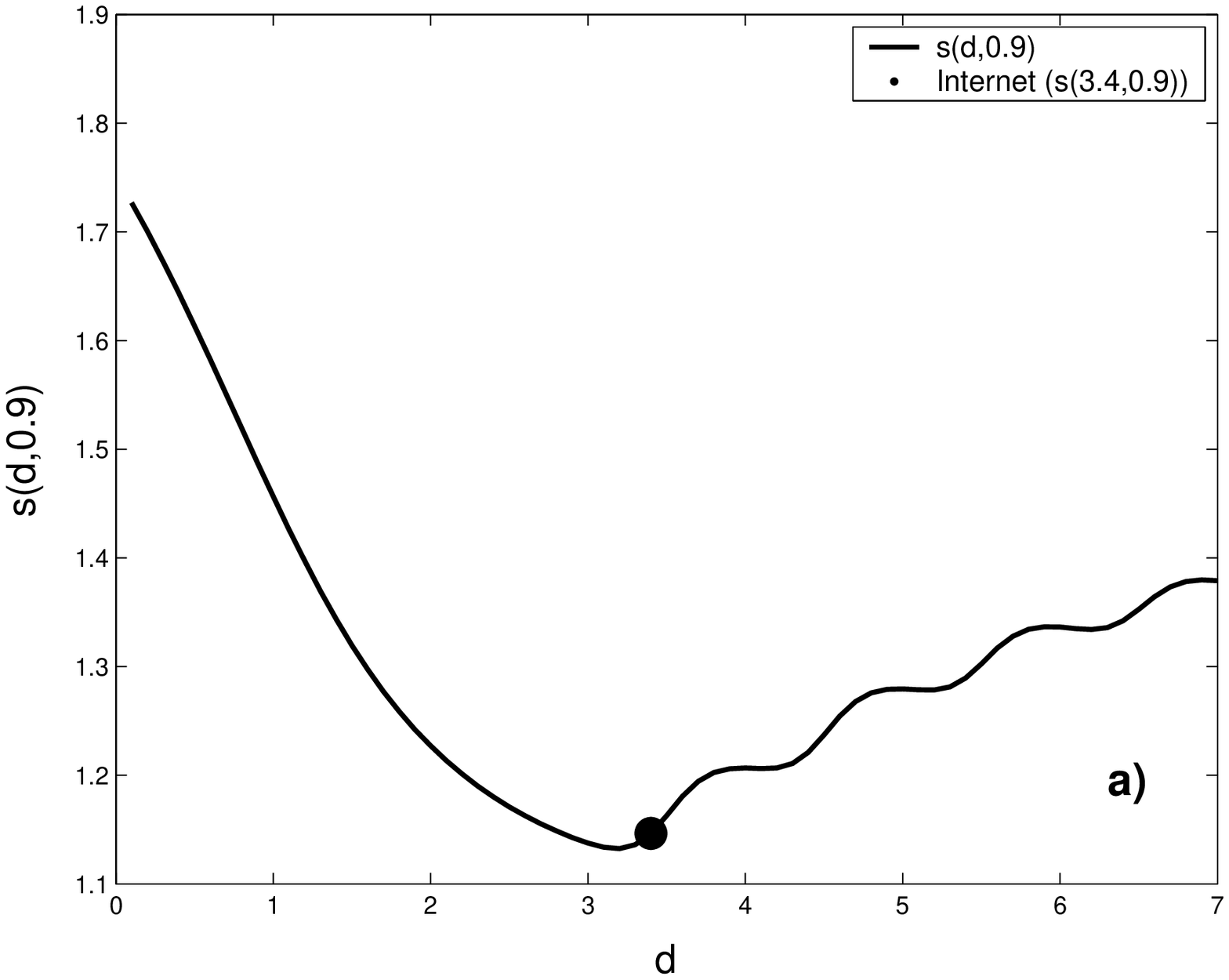,width=2in} \hspace*{0.5in}
  \epsfig{file=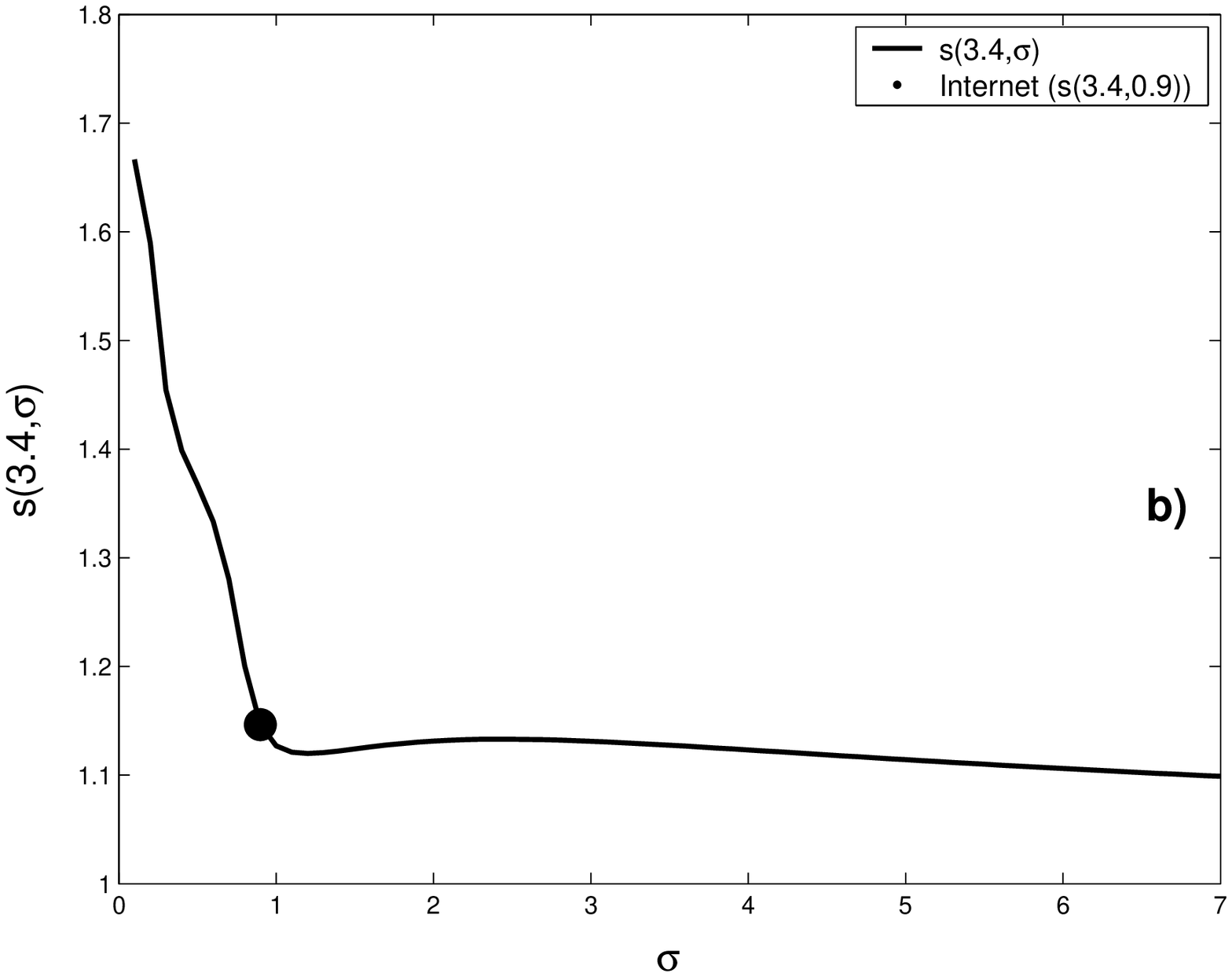,width=2in}}

%\hspace*{0.2in}

\centerline{\epsfig{file=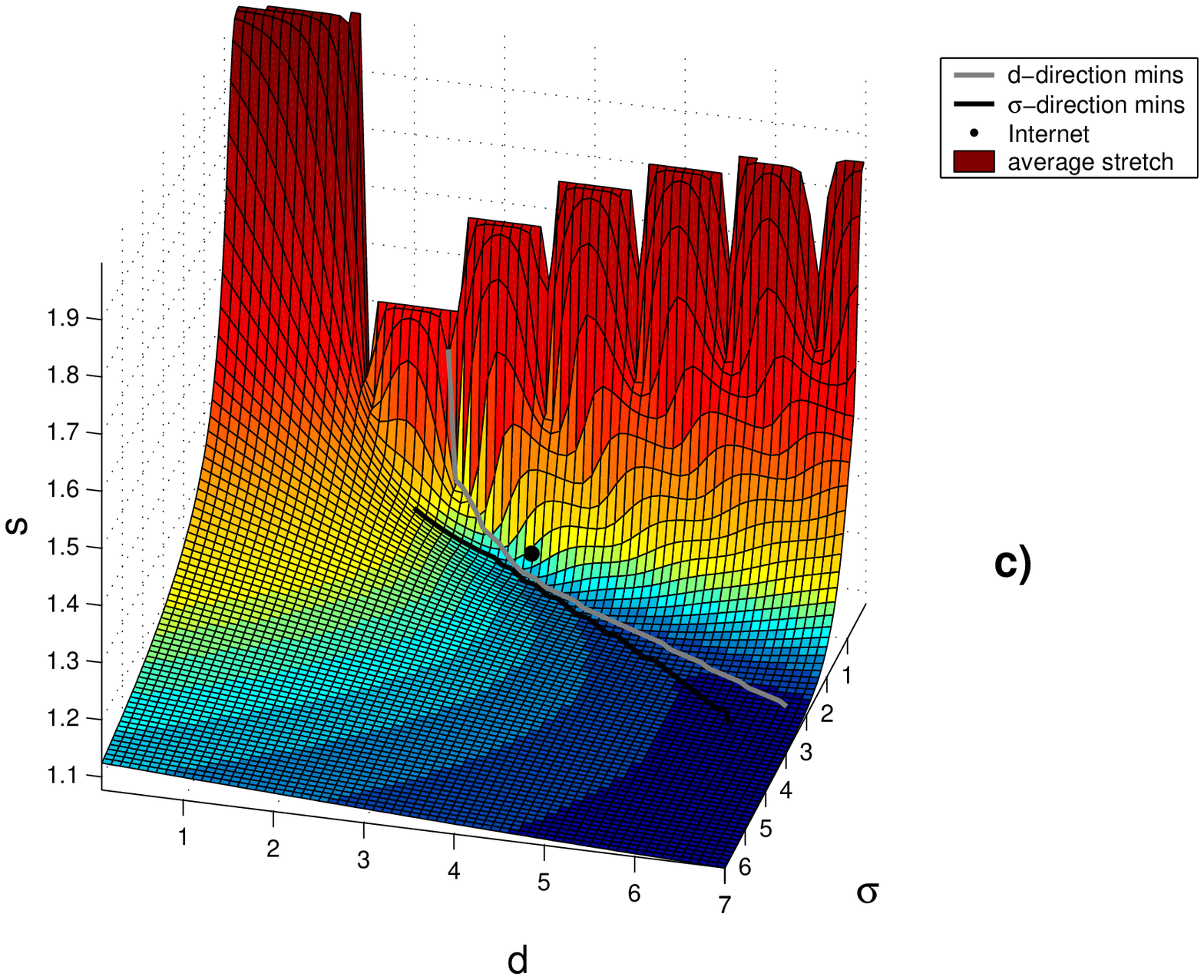,width=3.75in}}

%\hspace*{0.2in}

\centerline{\epsfig{file=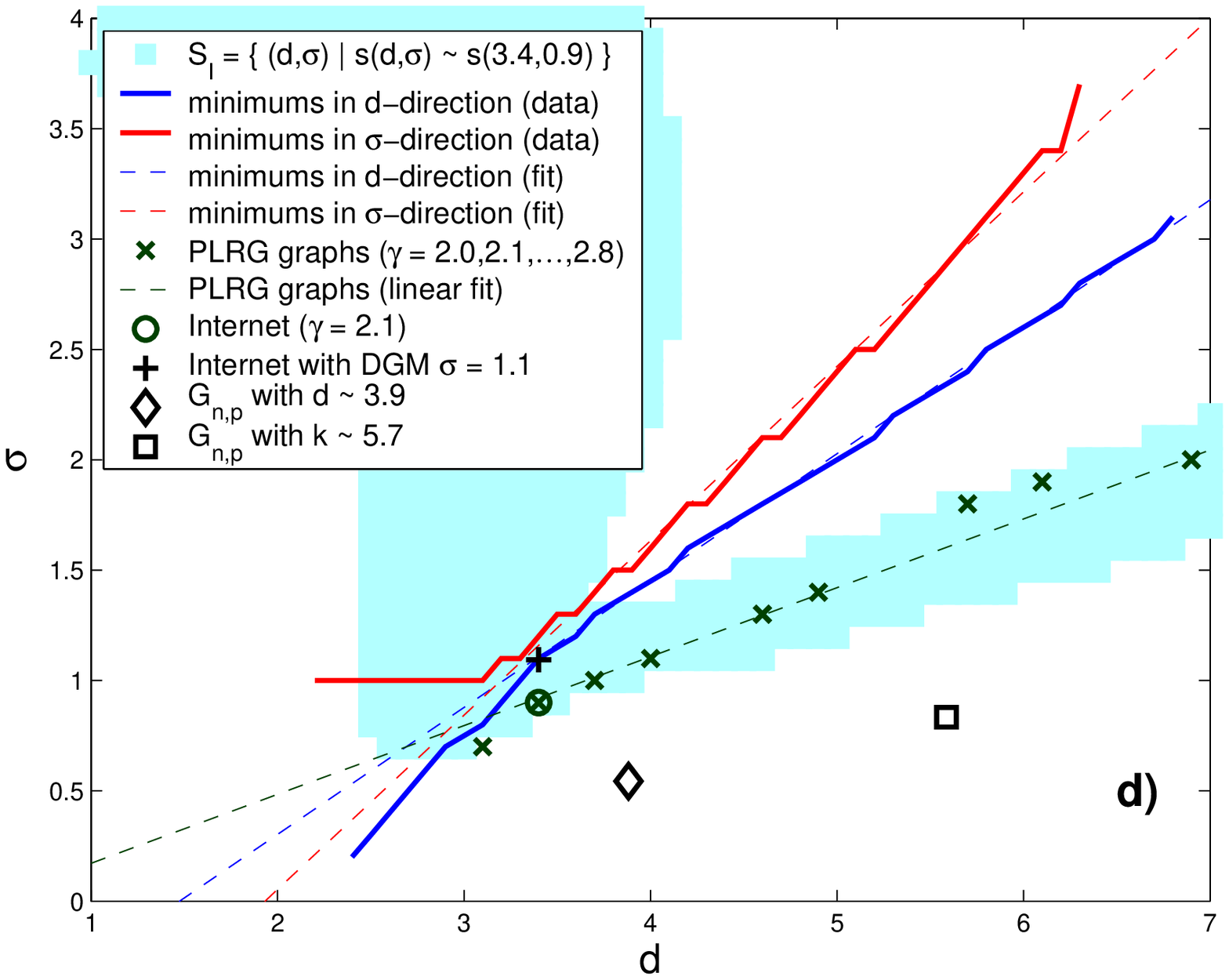,width=3in} \hspace*{0.5in}
\epsfig{file=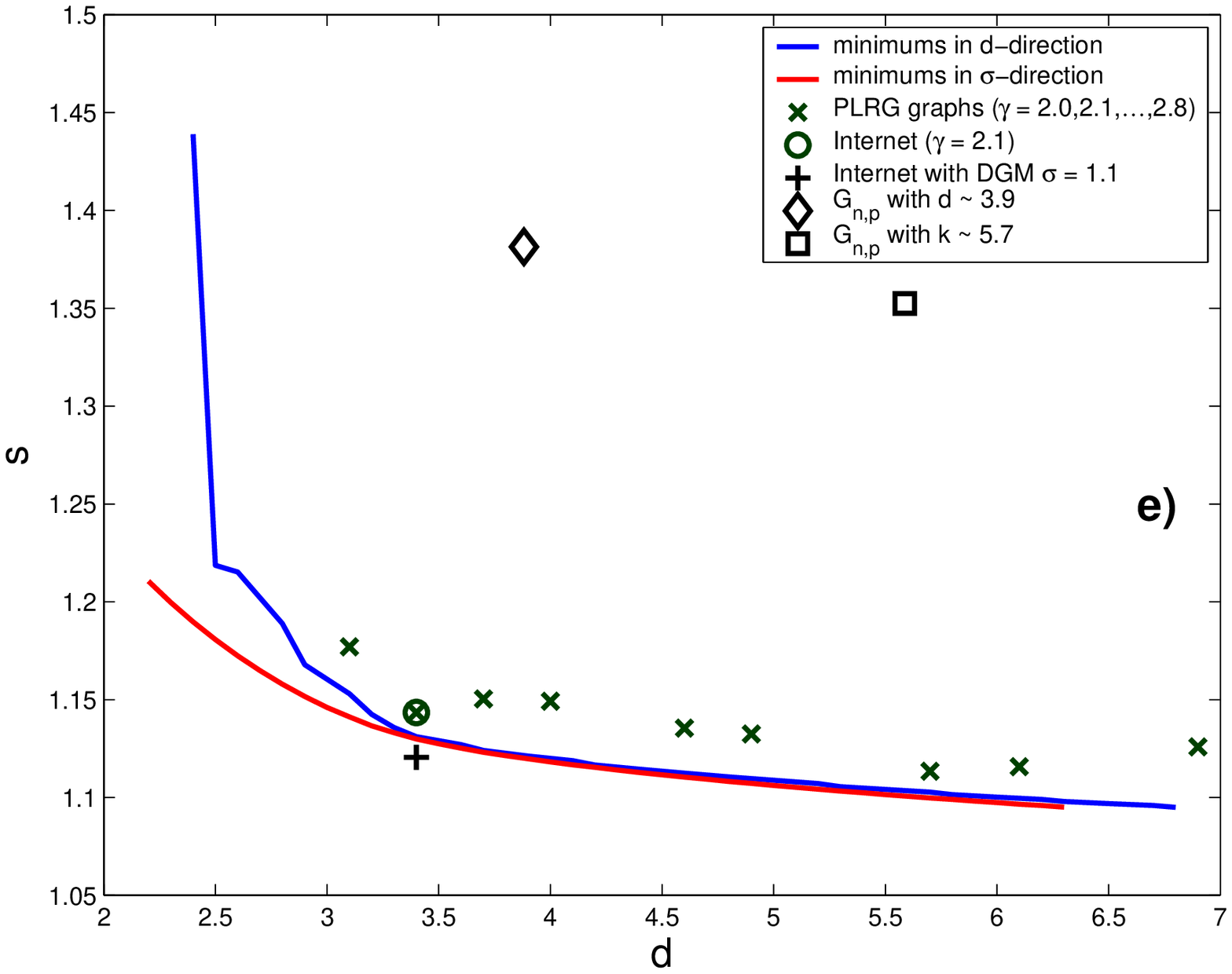,width=3in}}

\caption{\textbf{a),b)} The average stretch as functions of
$\overline{d}$ with $\sigma=0.9$ and of $\sigma$ with
$\overline{d}=3.4$ respectively. The Internet is represented by
the black dot. \textbf{c)} The average stretch as a function of
$\overline{d}$ and $\sigma$. The Internet is represented by the
black dot. The stretch minimums along the $\overline{d}$- and
$\sigma$-axes, $\sigma_{\overline{d}}^\ast(\overline{d}^\ast)$ and
$\sigma_{\sigma}^\ast(\overline{d}^\ast)$, are the grey and black
lines respectively. \textbf{d)} The projection of (c) onto the
$\overline{d}$-$\sigma$ plane. The solid blue (bottom) and red
(top) lines represent respectively
$\sigma_{\overline{d}}^\ast(\overline{d}^\ast)$ and
$\sigma_{\sigma}^\ast(\overline{d}^\ast)$ (the grey and black
lines from (c)). The dashed blue and red lines are their linear
fits in the {\em MSR}. The green crosses are the same as in
Fig.~\ref{fig:dd}(b), the green dashed line being their linear
fit. The Internet, $\gamma=2.1$, is circled. The shaded cyan area
is $S_I$ from the text. The black plus is the point with the
average distance observed in the Internet and the Gaussian width
predicted by the DGM model, $\overline{d}=3.4$, $\sigma=1.1$. The
black diamond and square are the distance distributions of the
${\cal G}_{n,p}$ graphs from Table~\ref{table:gnp} matching the
Internet average distance and node degree. \textbf{e)} The
projection of (c) onto the $\overline{d}$-$\overline{s}$ plane.
The notations are the same as in (d). The graph sizes $n\sim10^4$
everywhere.}

\label{fig:minimum}
\end{figure*}

The results in the previous sections suggest that the average
stretch depends more strongly on the characteristics of the graph
distance distribution (on its first and second moments, in
particular) than on the graph size. More specifically, we are
taking the distance distribution in a graph to be Gaussian,
(\ref{gaussian-pdf-norm}), and, hence, the average TZ stretch
$\overline{s}$ in (\ref{stretch-average}) is a function of the
average distance~$\overline{d}$ and the width of the distance
distribution~$\sigma$ in a graph,
$\overline{s}\equiv\overline{s}(\overline{d},\sigma)$. At this
point, we wish to explore the analytical structure of
$\overline{s}(\overline{d},\sigma)$ in more detail.

The natural starting point is to fix either $\overline{d}$ or
$\sigma$ in $\overline{s}(\overline{d},\sigma)$ to their observed
values, $3.4$ and $0.9$ respectively, and to consider two
functions, $\overline{s}(\overline{d},0.9)$ and
$\overline{s}(3.4,\sigma)$, which are shown in
Fig.~\ref{fig:minimum}(a,b). To our great surprise, we discover
that these two functions have {\em unique minimums\/} and that the
points corresponding to the Internet distance distribution (or,
simply, the {\em Internet points}) are {\em very close\/} to them.
In other words, one may get an impression that the Internet
topology has been carefully crafted to result in a distance
distribution that would minimize the average TZ stretch. Of
course, this can be only an impression and not an explanation
since the Internet evolution has had nothing to do with stretch.

The next question we have to ask is if the minimums we observe in
Fig.~\ref{fig:minimum}(a,b) correspond to a true local minimum of
$\overline{s}(\overline{d},\sigma)$. Our analytical results allow
us to collect enough data to draw Fig.~\ref{fig:minimum}(c), where
the stretch function $\overline{s}(\overline{d},\sigma)$ is shown
for $\overline{d},\sigma\in\left[0,7\right]$. Note that not all
regions of $(\overline{d},\sigma)$ correspond to Gaussian-like
distance distributions. Indeed, when $\sigma\gtrsim\overline{d}$,
distribution $f(d)$ from (\ref{gaussian-pdf-norm}) looks more like
an exponential decay since it is cut off from the left by
condition $d\geq1$. Also, when $\sigma$ is very small and the
continuous form of the Gaussian distribution approaches a
$\delta$-function, its discrete form (\ref{gaussian-pdf-norm})
goes to either a constant,
$f(d)\to\delta_{d,\left[\overline{d}\right]}$, when
$\overline{d}\neq(2k+1)/2$, or to a sum of two constants with
equal weights, $f(d)\to1/2(\delta_{d,k}+\delta_{d,k+1})$, when
$\overline{d}=(2k+1)/2$. This explains the peculiar peak formation
in the $\sigma\sim0$ area in the picture (see also
Appendix~\ref{sec:minimum_analysis}). Networks with such distance
distributions are easily constructible (complete networks, stars,
various forms of their interconnections, etc.). They all have
regular structure, and the exact knowledge of their structure is
required for precise stretch calculations. This is why our
analytical approach is slightly off in giving the precise answer
for stars ($\overline{s}=1.5$ instead of 1 for
$\overline{d}\sim2$, $\sigma\sim0$). Note, however, that for the
single case when $\sigma$ is allowed to be strictly 0, that is,
for the complete network case, $\overline{d}=1$, $\sigma=0$, we
obtain the correct answer for the average stretch, 2.

The $(\overline{d},\sigma)$-region for distance distributions in
realistic networks is, thus,
$0\lesssim\sigma\lesssim\overline{d}$, where we observe a concave
area in a form of channel, Fig.~\ref{fig:minimum}(c). The area is
characterized by particularly low stretch values, which makes us
call it the {\em minimal stretch region\/} (the MSR). The width
and depth of the MSR slowly {\em increase\/} as
$(\overline{d},\sigma)$ grow. In the area of smaller
$(\overline{d},\sigma)$, the MSR has a unique critical point,
which we call the {\em MSR apex}. The Internet point is located
very close to the apex, which is characterized by the shortest
distance between the sets of minimums of $\overline{s}$---along
the $\overline{d}$- and $\sigma$-axes. We may express these sets
as two functions, which we denote as
$\sigma_{\overline{d}}^\ast(\overline{d}^\ast) = \bigm\{
(\sigma,\overline{d}) \bigm| \partial\overline{s} /
\partial\overline{d} = 0 \bigm\}$ and $\sigma_\sigma^\ast(\overline{d}^\ast)
= \bigm\{ (\sigma,\overline{d}) \bigm| \partial\overline{s} /
\partial\sigma = 0 \bigm\}$ respectively. We find that
$\sigma_{\overline{d}}^\ast(\overline{d}^\ast)$ and
$\sigma_{\sigma}^\ast(\overline{d}^\ast)$ {\em almost touch\/}
each other at the apex. Since the intersection of these two
functions would correspond to a stationary point of
$\overline{s}(\overline{d},\sigma)$, we call the apex a {\em
quasi-stationary\/} point emphasizing that the both derivatives of
$\overline{s}(\overline{d},\sigma)$ are {\em nearly} zero at this
point.

The apex can be more easily observed in Fig.~\ref{fig:minimum}(d)
showing a projection of Fig.~\ref{fig:minimum}(c) on the
$\overline{d}$-$\sigma$ plane. The solid lines representing the
above two sets of minimums forming the MSR, almost touch each
other near the apex, and the Internet point is very near their
closest segment.

An opportunity to look at the apex from yet another angle is
presented in Fig.~\ref{fig:minimum}(e) showing a projection of
Fig.~\ref{fig:minimum}(c) on the $\overline{d}$-$\overline{s}$
plane. We see that starting from the apex, as~$\overline{d}$
increases, the stretch values along
$\sigma_{\overline{d}}^\ast(\overline{d}^\ast)$ and
$\sigma_{\sigma}^\ast(\overline{d}^\ast)$ become virtually equal
and slowly {\em decrease\/} as the average distance grows. We also
note that ${\cal G}_{n,p}$ graphs are far away from the apex and
that they have average stretch values that are far from minimal.

We can see now that the apex is indeed a critical or ``phase
transition'' point since it is located at the boundary of the two
regions of the average stretch function. The first region, the
MSR, is characterized by lowest possible stretch values
corresponding to distance distributions observed in real-world
graphs. The second region, with substantially higher average
stretch values, corresponds to distance distributions in more
regular graphs.

To illustrate this point in more detail, we turn our attention
back to Fig.~\ref{fig:minimum}(d). We see that the two sets of
minimums, $\sigma_{\overline{d}}^\ast(\overline{d}^\ast)$ and
$\sigma_{\sigma}^\ast(\overline{d}^\ast)$, are linear in the MSR
with sufficiently large $\sigma$, $\sigma\gtrsim1$, where the
continuous approximation of the distance distribution works
particularly well. The two top dashed lines in
Fig.~\ref{fig:minimum}(d) represent the linear fits of
$\sigma_{\overline{d}}^\ast(\overline{d}^\ast)$ and
$\sigma_{\sigma}^\ast(\overline{d}^\ast)$ in the area with
$\sigma\gtrsim1$. The exact location of the intersection of these
fits is $(\overline{d}^\star,\sigma^\star)=(3.16,0.97)$, while the
two closest points on the data curves for
$\sigma_{\overline{d}}^\ast(\overline{d}^\ast)$ and
$\sigma_{\sigma}^\ast(\overline{d}^\ast)$ are
$\sigma_{\overline{d}}^\ast(3.59)=1.20$ and
$\sigma_{\sigma}^\ast(3.55)=1.29$. If the linear form of
$\sigma_{\overline{d}}^\ast(\overline{d}^\ast)$ and
$\sigma_{\sigma}^\ast(\overline{d}^\ast)$ sustained in the area
with smaller $\sigma$ as well, then
$\sigma_{\overline{d}}^\ast(\overline{d}^\ast)$ and
$\sigma_{\sigma}^\ast(\overline{d}^\ast)$ would intersect at
$(\overline{d}^\star,\sigma^\star)$, where we would observe a true
stationary point of $\overline{s}(\overline{d},\sigma)$, which we
could then test for the presence of an extremum of the stretch
function. This does not happen, however. Instead, as
$\overline{d}$ and $\sigma$ become small, the linear behavior
breaks near the apex due to increasingly ``more discrete''
structure of the distance distribution. See more on this in
Appendix~\ref{sec:minimum_analysis}.

In Appendix \ref{sec:minimum_analysis}, we show that linearity of
$\sigma_{\overline{d}}^\ast(\overline{d}^\ast)$ and
$\sigma_{\sigma}^\ast(\overline{d}^\ast)$ can be analytically
derived from the fact that the distance distribution is taken to
be Gaussian. Of course, this does not explain why the {\em
Internet\/} point is so close either to the MSR or to its apex.

The linear form of $\sigma_{\overline{d}}^\ast(\overline{d}^\ast)$
and $\sigma_{\sigma}^\ast(\overline{d}^\ast)$ sheds some light on
a closely related issue of why the average stretch is virtually
independent of $\gamma$. In Fig.~\ref{fig:minimum}(d), the shaded
area represents a set of $(\overline{d},\sigma)$, for which the
average stretch is approximately the same as for the Internet,
$S_I = \bigm\{ (\overline{d},\sigma) \bigm |
\overline{s}(\overline{d},\sigma) \sim \overline{s}(3.4,0.9)
\bigm\}$. In other words, it is a projection of the cyan area in
Fig.~\ref{fig:minimum}(c) on the $\overline{d}$-$\sigma$ plane. We
see that in the MSR, the $S_I$ boundaries are almost parallel
straight lines. Therefore, if the average stretch is to be
independent of $\gamma$, which is observed in
Section~\ref{sec:stretch_distribution}, then the points
representing distance distributions in power-law graphs,
$(\overline{d}_\gamma,\sigma_\gamma)$, from Fig.~\ref{fig:dd}(b)
should lie along the $S_I$ boundaries, and this is what indeed
happens. Yet again, the linear relation between
$\overline{d}_\gamma$ and $\sigma_\gamma$ in the power-law graphs,
and the fact that this relation is just as required for the
average TZ stretch being virtually independent of $\gamma$, come
from two seemingly disjoint domains.

To finish the list of various ``coincidences,'' we construct a
linear fit of ($\overline{d}_\gamma,\sigma_\gamma$) (the
bottom-most dashed line in Fig.~\ref{fig:minimum}(d)). The
Internet point, $\gamma=2.1$, lies on this line. Our numeric
analysis shows that the Internet value of $\gamma=2.1$ is a unique
value of $\gamma$ minimizing the distance between the linear fit
of ($\overline{d}_\gamma,\sigma_\gamma$) and
$(\overline{d}^\star,\sigma^\star)$, which is the intersection of
the linear fits of $\sigma_{\overline{d}}^\ast(\overline{d}^\ast)$
and $\sigma_{\sigma}^\ast(\overline{d}^\ast)$. In other words, the
Internet distance distribution is the point that is {\em
closest\/} to the MSR apex, compared to distance distributions in
all other scale-free graphs with power-law node degree
distributions.

\section{Conclusions}\label{sec:conclusions}

Of course, the TZ scheme, reducing, in principle, the routing
table size to about $50$ entries for $10^4$-node scale-free
networks, and making routing decision running time constant,
cannot pretend to be a realistic Internet interdomain routing
scheme. First, it is static. Second, addressing in interdomain
routing is based on IP addresses rather than on interdomain graph
node labels, that is, AS numbers.\footnote{Although there are some
proposals, {\em Atomized Routing},
\cite{BroCla01b,BroCla01c,VerBroCla03}, and {\em ISLAY},
\cite{islay}, suggesting to ``fix'' this. If this is ``fixed,'' we
should be ready to face the problem of accelerated rates of growth
of the total number of ASs.} Third, it assumes availability of the
global topology view.

Most importantly, in the context of our work, the scheme is not a
stretch-1 scheme. Indeed, interdomain routing in the Internet is
essentially shortest path routing.\footnote{Shortest path routing
in the Internet is perturbed by various administrative constraints
called {\em policies}, the routing protocol, BGP, being a policy
routing tool. For measurements of stretch produced by policy
routing, see \cite{TaGoSh01}.} A routing scheme that would prevent
a pair of ASs from utilizing a peering link between them is not
realistic, of course. Thus, any stretch $s>1$ routing scheme
applied to the Internet would involve augmentation, in one form or
another, of the routing information provided by the scheme with
the shortest path routing information for non-shortest paths. This
explains why we are concerned with the average stretch produced by
a scheme.

Our principal finding that the average TZ stretch on the Internet
graph is reasonably low opens a well-defined path for the future
work in the area of applying relevant theoretical results obtained
for routing to realistic scale-free networks (see the next
section). If the average stretch of even the ``exceptional'' TZ
scheme turned out to be relatively high, the scheme would be
inapplicable {\em in principle} (not just {\em in practice}),
which would essentially close the above path, demonstrate
impossibility to construct efficient and scalable routing for the
Internet, and call for searching one somewhere beyond the
traditional graph-theoretical approach.

As we mention in the introduction, our finding that the Internet
distance distribution is in a close neighborhood of the MSR apex
cannot be explained in the present idea set since the Internet, as
we know it today, has nothing to do with stretch. While we lack
sufficient information to show cause for this effect, we do
believe it strongly suggests the analytical structure of the
average stretch function may be an indirect (or even direct)
indicator of some yet-to-be discovered processes that have
influenced the Internet's topological evolution. In other words, a
rigorous explanation of this phenomenon would probably require
much deeper understanding of the Internet evolution principles
(that are far from being even known if we accept the critique of
the BA model and alike) and demonstration of a link between them
and the TZ scheme.

We believe that an explanation of this effect will most probably
have the following pattern. The Internet evolution principles turn
out to be such that they minimize $X$, where $X$ is some known or
yet unknown characteristic of a network. At the same time, the
distance to the MSR apex turns out to be a monotonically
increasing function of $X$. There are reasons to believe that the
distance to the apex is not something random. Indeed, as we have
seen, the apex is a {\em unique\/} critical point of the average
TZ stretch function, and, at the same time, the TZ scheme is also
``exceptional'' in a sense that it delivers a nearly optimal first
possibility to deviate from incompressible shortest path routing.
The outlined pattern would create a necessary link between the two
seemingly disjoint domains.

\section{Future work}\label{sec:future_work}

The list of immediate practical next questions that remain open
include:

\begin{itemize}
    \item What is the average stretch produced by the {\em
    Cowen\/} scheme on scale-free graphs?

    Obtaining analytical results for the Cowen scheme is a harder
    problem. However, one can expect that the average stretch
    would be lower. Indeed, it is easy to see that making high-betweenness
    nodes belong to the LS should decrease the average stretch.
    The Cowen LS construction algorithm uses the greedy set
    cover algorithm by Lov\'{a}sz preferring high degree nodes,
    \cite{lovasz75}, but as observed in \cite{VaPaVe02a},
    betweenness is linearly correlated with node degrees in the
    Internet. The expectation of the Cowen scheme producing an
    average stretch that is lower than in the TZ case is in a
    good agreement with the general stretch-space trade-off
    since the Cowen scheme has a higher local memory space
    upper bound.

    \item Do any routing schemes based on the {\em additive\/} stretch
    factor deliver lower average stretch on scale-free graphs?

    The multiplicative stretch factor may be too coarse for short
    distances that prevail in scale-free graphs.

    \item What is the memory space lower bound for {\em shortest path\/}
    routing on scale-free graphs?

    We know that generic shortest path routing is incompressible.
    However, the situation is better for almost all graphs. We do
    not know any bounds for scale-free graphs. Answering this
    question has very important practical implications since
    (policy-constrained) shortest path routing seems to be a
    requirement for Internet interdomain routing.

    \item More specifically, can any upper bounds be obtained for
    the total number of stretch $s>1$ paths produced by existing $s>1$
    routing schemes on scale-free graphs?

    If these upper bounds are found to be as low as the total space upper
    bounds, then the original routing information can be augmented with the
    $s=1$ information for $s>1$ paths without increasing the space upper bounds.

    \item What bounds can be obtained for {\em dynamic\/} routing
    on scale-free graphs?

    Of course, no realistic Internet routing can neglect the
    adaptation cost considerations. While of critical practical
    importance, obtaining various bounds for dynamic routing on
    scale-free graphs seems to be the hardest problem among those we
    list here. Furthermore, one can hardly expect such bounds to
    be low, taking into consideration rather pessimistic lower
    bounds obtained for dynamic routing on generic networks
    (cf.~Section~\ref{sec:routing}). In addition, it is clear that
    the TZ scheme cannot be easily modified to perform well in the
    dynamic case since the scheme labels nodes with
    topology-sensitive information. In other words, the scheme is
    not name-independent. As soon as topology changes, nodes need to
    be relabelled. Significant progress in construction of
    name-independent static low-stretch routing schemes has been
    recently made by Arias, Cowen, et al.~in~\cite{ArCoLaRaTa03}.

\end{itemize}

On the theoretical side, which may turn out to have practical
implications as well, the explanation of Internet distance
distribution proximity to the MSR apex appears to be a very
interesting problem. It would be much easier to solve, of course,
if the integral in~(\ref{stretch-superlong}) could be evaluated
analytically. What might be an alternative set of assumptions that
would make an expression analogous to~(\ref{stretch-superlong})
analytically solvable? Can similar results be obtained for other
forms of distance distributions, and, yet more importantly, for
other routing schemes (the Cowen scheme, or the stretch-5 routing
scheme with the $\tilde{O}(n^{1/2})$ local memory space upper
bound obtained in \cite{EiGaPe03}, for example)? A technical issue
with {\em experimental\/} (vs.\ analytical) studies of the MSR is
that we do not know an efficient algorithm to generate graphs with
a given distance distribution.

In any case, the explanation of Internet's proximity to the apex
is of great theoretical interest, as the fundamental laws
governing the Internet evolution remain unclear. Therefore, on the
practical side, a proper explanation of this effect may help us,
for example, in out intent to move, \cite{ChaJamWil03}, from
purely {\em descriptive} Internet evolution models to more {\em
explanatory} ones, in the terminology of the program outlined by
the authors of~\cite{WilCriticality}.

The Internet started as a small research network designed and
fully controlled by a group of few
enthusiasts~(\cite{internet-history}). Today, after a series of
``phase transitions,'' it has evolved to a huge network
interconnecting tens of thousands of independent and even
adversarial networks without a single point of external control.
This makes the Internet a ``self-governing,'' ``self-evolving''
complex system, a research subject of areas of physics
(statistical mechanics, in particular) studying evolution of such
systems in general. Construction of realistic, efficient, and
scalable routing for this ``new'' Internet is an interesting and
challenging task lying ahead of us.

\newpage
\appendix
\noindent{\LARGE\bfseries Appendices}

\section{The Kleinrock-Kamoun stretch on the Internet}\label{sec:kk_stretch}

In this appendix, we calculate a rough estimate of the stretch
factor of the Kleinrock-Kamoun (KK) hierarchical routing scheme,
\cite{KK77}, applied to the observed Internet interdomain
topology. We find that the stretch is very high, which is
consistent with the observation made in \cite{KK77} that the
approach used there works reasonably well only for sparsely
connected networks. The scale-free networks, on the contrary, are
extremely densely connected.

Recall that \cite{KK77} assumes existence of a hierarchical
partitioning of a network of size $n$ into $m$ levels of clusters.
Each $k$-level cluster consists of $n^{1/m}$ $(k-1)$-level
clusters, $k = 1 \ldots m$, $0$-level clusters being nodes. The
optimal clustering is achieved when $m \sim \log n$. There are few
other fairly strong assumptions about the properties of required
partitioning. Neither algorithm for its construction nor proof of
its existence are delivered, but if it does exist then the stretch
factor is shown to be
\begin{equation}\label{kk_stretch:input}
    s = 1 + \frac{1}{\overline{d}}
    \sum_{k=1}^{m-1}\left[1-\frac{n^{\frac{k}{m}}-1}{n-1}\right]d_k,
\end{equation}
where $\overline{d}$ is the network average distance and $d_k$ is
the diameter of a $k$-level cluster.

It is further assumed in \cite{KK77} that both the network
diameter and average distance are power-law functions of the
network size. This is certainly not true for scale-free networks
with power-law {\em node degree\/} distributions. For the very
recent results on the average distance in such networks see the
references mentioned in Section \ref{sec:scale-free_networks}. In
the numerical evaluations in this appendix, we use the value of
$\overline{d} \sim 3.6$ observed in the Internet.

Much fewer results are available for the network diameter.
However, in \cite{ChLu03}, it is shown that the diameter of
networks with power-law node degree distribution with exponent
$\gamma$ lying between $2$ and $3$ scales almost surely as
$\Theta(\log n)$. For the Internet, $\gamma \sim 2.1$, and since
the Internet size $n \sim 1.5 \times 10^4$ is relatively large, we
may write the Internet diameter $D$ as $D \sim c \log n$ with some
multiplicative coefficient $c$. The observed value of $D$, $D \sim
13$ (\cite{huston-bgp-site}), defines $c$ then.

The size of a $k$-level cluster is obviously $n^{k/m}$ but nothing
rigorous can be said about its degree distribution since there is
no procedure for its construction. Thus, it is natural to assume
that its degree distribution is also power-law with $2<\gamma<3$,
which gives an estimate of the $k$-level cluster diameter as $d_k
\sim c \log n^{k/m} \sim Dk/m$. Substituting this in
(\ref{kk_stretch:input}) and performing summation gives
\begin{equation}\label{kk_stretch:exact_output}
    s \sim 1 + \frac{D}{2 \overline{d}} \left[
    m \frac{n}{n-1} -
    \frac{n (n^{\frac{2}{m}} - 1)} {(n-1) (n^\frac{1}{m} - 1)^2} +
    \frac{2}{m} \frac{n^\frac{1}{m}}{(n^\frac{1}{m}-1)^2} \right] .
\end{equation}
Using the numerical values for $n$, $\overline{d}$, $D$, and
optimal $m=10$, we can see that the KK stretch factor on the
Internet interdomain topology is
\begin{equation}
    s \sim 15.
\end{equation}
Note that a 15-time path length increase in the Internet would
lead to AS path lengths of $\sim 55$ and IP hop path lengths of
$\sim 150$.

\begin{figure}
  \centerline{\epsfig{file=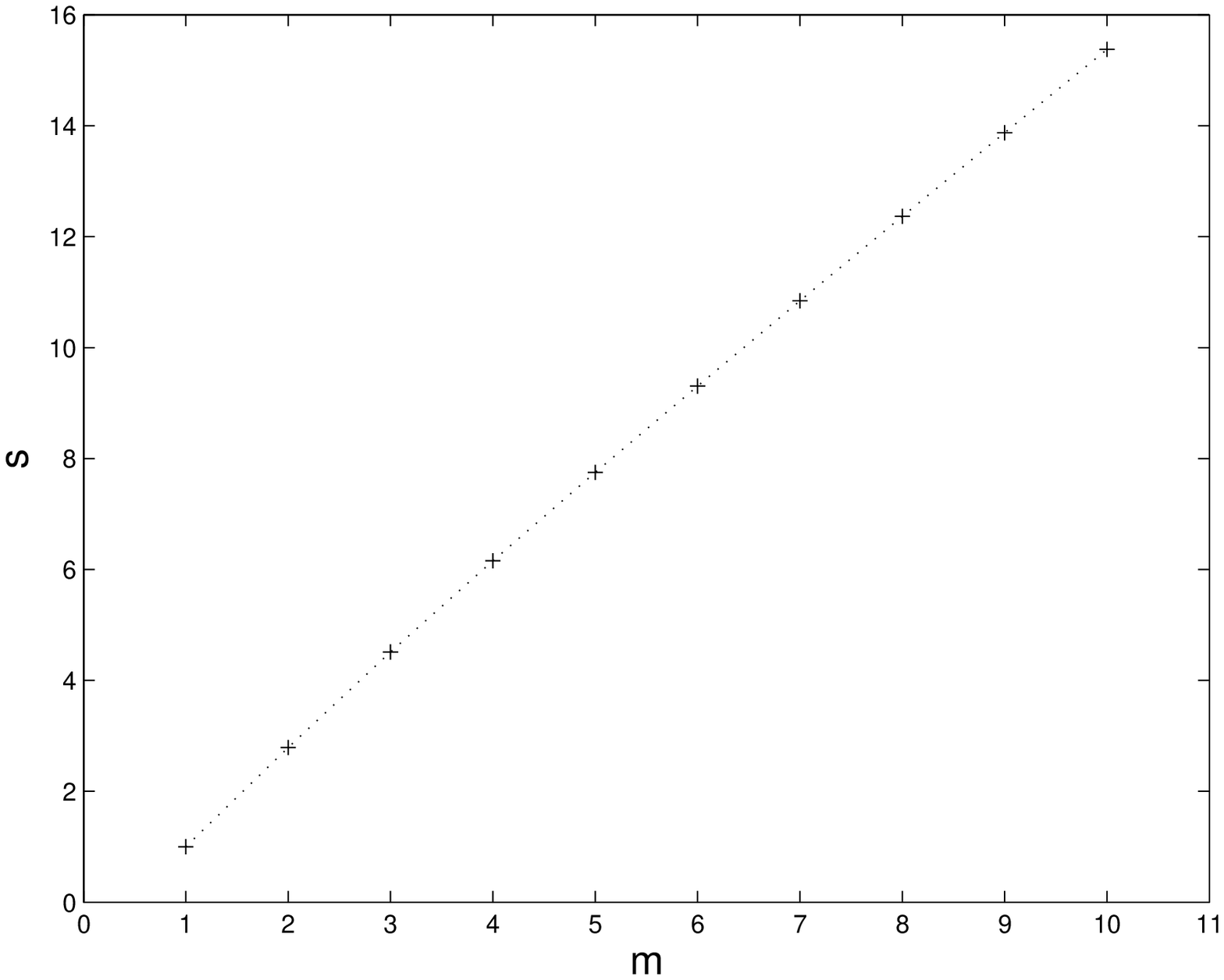,width=3in}}
  \caption{The KK stretch factor $s$ as a function
  of the number of levels of hierarchy $m$.}
  \label{fig:kk_stretch}
\end{figure}
Stretch factors for smaller, non-optimal values of $m$ are shown
in Fig.\ \ref{fig:kk_stretch}. The fact that the stretch grows
almost linearly with the number of hierarchical levels follows
directly from (\ref{kk_stretch:exact_output}), which, for large
$n$, can be rewritten as
\begin{equation}
    s \sim 1 + \frac{D}{2 \overline{d}}(m - 1).
\end{equation}
Using $D \sim \log n$, $\overline{d} \sim \log \log n$
(\cite{ChLu03}), and optimal $m \sim \log n$, we obtain the
following estimate of the stretch factor as a function of the
network size:
\begin{equation}
    s \sim \frac{\log^2 n}{\log \log n}.
\end{equation}

\section{Proofs}\label{sec:proofs}

In this appendix, we prove various statements made in Section
\ref{sec:analytical_results}.

\subsection{Claims \ref{claim:no_difference} and
\ref{claim:gi}}\label{sec:proofs:gi}

A rigorous proof would involve the same type of argument as used
to show that the hypergeometric distribution converges to the
binomial distribution for small selections from large sets. We
provide a close outline of the proof.

Suppose we have a set of $n$ objects of two types: $n_1$ objects
of type 1 and $n_2$ objects of type 2, $n_1+n_2=n$ and $\vartheta
= n_1/n$. Recall that the hypergeometric distribution,
\begin{equation}\label{hypergeometric}
    p_h(x) = \frac{\binom{n_1}{x}\binom{n_2}{m-x}}{\binom{n}{m}},
\end{equation}
gives the probability that a random sample of $m$ objects from the
set of all objects contains $x$ objects of type 1. If sampling is
with replacement---that is, as soon as one object is selected, it
is immediately returned back to the set before the next object is
selected,---then the probability that $x$ type-1 objects have been
selected after $m$ one-object selections is given by the binomial
distribution,
\begin{equation}\label{binomial}
    p_b(x) = \binom{m}{x} \vartheta^x (1-\vartheta)^{m-x}.
\end{equation}
One can easily see that probabilities for small selections with
and without replacement converge in the limit of large $n$,
\begin{equation}\label{hyper-to-binom}
    p_h(x) \xrightarrow
    [\begin{subarray}{c} m/n \to 0 \\
    n \to \infty \end{subarray}]{} p_b(x).
\end{equation}

Suppose now that $f(d)$ and $F(d)$ are distance p.d.f.\ and
c.d.f.\ of some graph of size $n$, diameter $D$, and that the
graph LS $L$ is of size $q$. The problem of finding the
probability $g_i(d)$ that the $i$'th closest landmark is at
distance $d$ is equivalent to the following object selection
problem. The total number of objects to select from is $n$, the
number of types of objects is $D$, and the number of objects of
type $d$, $d = 0 \ldots D$, is $f(d)n$. In addition, $q$ random
objects are marked as landmarks. Since landmarks are randomly
marked, $g_i(d)$ is the probability that a randomly selected
object is of type $d$, which is $f(d)$ by definition, times the
probability that $i-1$ random objects are of (possibly different)
types $d_- \leq d$ (corresponding to the closer landmarks), times
the probability that $q-i$ random objects are of (possibly
different) types $d_+ \geq d$ (corresponding to the farther
landmarks). That is, denoting the latter two probabilities by
$p_-(d)$ and $p_+(d)$ respectively, we can write that
\begin{equation}\label{gip-p+}
    g_i(d) = c_i p_-(d) f(d) p_+(d)
\end{equation}
with some normalization coefficient $c_i$.

The probability than one random object is of type $d_-$ ($d_+$) is
$F(d)$ ($1-F(d)$) by definition, but obtaining the exact answer
for $p_-(d)$ ($p_+(d)$) involves some combinatorics resulting in a
generalization of the hypergeometric distribution
(\ref{hypergeometric}), which is left as an exercise for a
rigorous reader. However, since $q/n\xrightarrow[n\to\infty]{}0$,
sampling without replacement can be approximated by sampling with
replacement, cf.\ (\ref{hyper-to-binom}), which leads to the
following approximations for $p_-(d)$ and $p_+(d)$:
\begin{eqnarray}
  p_-(d) &\to& F(d)^{i-1}, \\
  p_+(d) &\to& (1-F(d))^{q-i}.
\end{eqnarray}
The analogy to the order statistics becomes straightforward now.

The normalization coefficient~$c_i$ in~(\ref{gip-p+}) is defined
by the normalization condition
\begin{equation}
    1 = \sum_{d=0}^D g_i(d) =
    c_i \sum_{d=0}^D F(d)^{i-1} f(d) (1-F(d))^{q-i},
\end{equation}
which can be expressed as the following Lebesgue-Stieltjes
integral:
\begin{equation}
    1 = c_i \int\limits_0^D F(d)^{i-1}
    (1-F(d))^{q-i} \, \dif F(d)
    = c_i \int\limits_0^1 F^{i-1}
    (1-F)^{q-i} \, \dif F.
\end{equation}
In the continuous limit, evaluation of the corresponding Riemann
integral gives
\begin{equation}
    1 = c_i
    \frac{\Gamma(i)\Gamma(q-i+1)}{\Gamma(q+1)} =
    c_i \left[ i \binom{q}{i} \right]^{-1}.
\end{equation}

\subsection{Claim \ref{claim:average-cluster}}\label{sec:proofs:average-cluster}

If we denote by $\tilde{G}_1(d)$ the c.c.d.f.\ for the distance to
the closest landmark,
\begin{equation}\label{G1-sum}
    \tilde{G}_1(d) = \sum_{\tilde{d}=d}^D g_1(\tilde{d}),
\end{equation}
then according to the cluster definition (\ref{cluster_def}), the
average cluster size is
\begin{equation}\label{average-C-sum}
    \overline{|C|} = \sum_{d=0}^D n f(d)
    \tilde{G}_1(d).
\end{equation}

Using the explicit expression for~$g_1$ in~(\ref{gi}), we can
calculate an upper bound for~$\tilde{G}_1$ by representing the sum
in~(\ref{G1-sum}) as the following Lebesgue-Stieltjes integral:
\begin{equation}
    \tilde{G}_1(d) = q \int\limits_d^D
    (1-F(\tilde{d}))^{q-1} \, \dif F(\tilde{d})
    = q \int\limits_{F(d)}^1
    (1-F)^{q-1} \, \dif F.
\end{equation}
Evaluation of the corresponding Riemann integral gives
\begin{equation}\label{G1-final}
    \tilde{G}_1(d) \leq (1-F(d))^q,
\end{equation}
where equality is attained in the continuous limit. Note that the
obtained upper bound, $(1-F(d))^q$, is equivalent to the c.c.d.f.\
for the geometric distribution, $p_g(x) =
\vartheta(1-\vartheta)^x$, with the success probability $\vartheta
= F(d)$ and number of trials $x=q-1$. The reason for that becomes
transparent after formulating the problem in the object selection
framework considered in Appendix \ref{sec:proofs:gi}.

Substituting (\ref{G1-final}) in (\ref{average-C-sum}) completes
the proof:
\begin{equation}
    \overline{|C|} =
    n \int\limits_0^D \tilde{G}_1(d) \, \dif F(d) \leq
    n \int\limits_0^1 (1-F)^q \, \dif F \leq
    \frac{n}{q+1}.
\end{equation}

Note that clusters are originally defined as objects ``inverse''
to balls \mbox{$B(v)= \bigm\{ b\ \in V \bigm| \delta(b,v) <
\delta(v,L(v)) \bigm\}$}. They are ``inverse'' in the following
sense: \mbox{$v \in B(w) \Leftrightarrow w \in C(v)$}. If we
denote the c.d.f.\ of $g_1(d)$ by $G_1(d)$, then the average ball
size is
\begin{equation}
    \overline{|B|}=\sum_{n=0}^D n g_1(d) F(d) =
    n \int\limits_0^D F(d) \, \dif G_1(d).
\end{equation}
Since (\ref{average-C-sum}) can be written as
\begin{equation}
    \overline{|C|} = n \left( 1 -
    \int\limits_0^D G_1(d) \, \dif F(d)\right),
\end{equation}
the known fact that the average ball and cluster sizes are equal
is just a consequence of integration by parts of the following
Lebesgue-Stieltjes integral:
\begin{equation}
    1 = F(d)G_1(d)\bigg|_0^D =
    \int\limits_0^D F(d) \, \dif G_1(d) +
    \int\limits_0^D G_1(d) \, \dif F(d).
\end{equation}

\subsection{Claim \ref{claim:shortcut}}\label{sec:proofs:shortcut}

Note that the first case in (\ref{stretchval-shortcut}), $z<y$,
accounts exactly for stretch-1 paths from $w$ to destinations $v
\in C(w)$, (see definition (\ref{cluster_def})).

The second case, $z<x$, accounts for the degenerate shortcut on
paths from $w$ to $L(v)$ that go through $v$. On such paths, $x =
z + y$, the length of the shortcut portion of the total path from
$w$ to $v$ is zero, and the stretch is 1. It is easy to see that,
on small-world graphs, paths from $w$ to $L(v)$ that do not go
through $v$ but that still have shortcuts are rare and hard to
account for without knowing the graph topology.

\begin{figure}
  \centerline{\epsfig{file=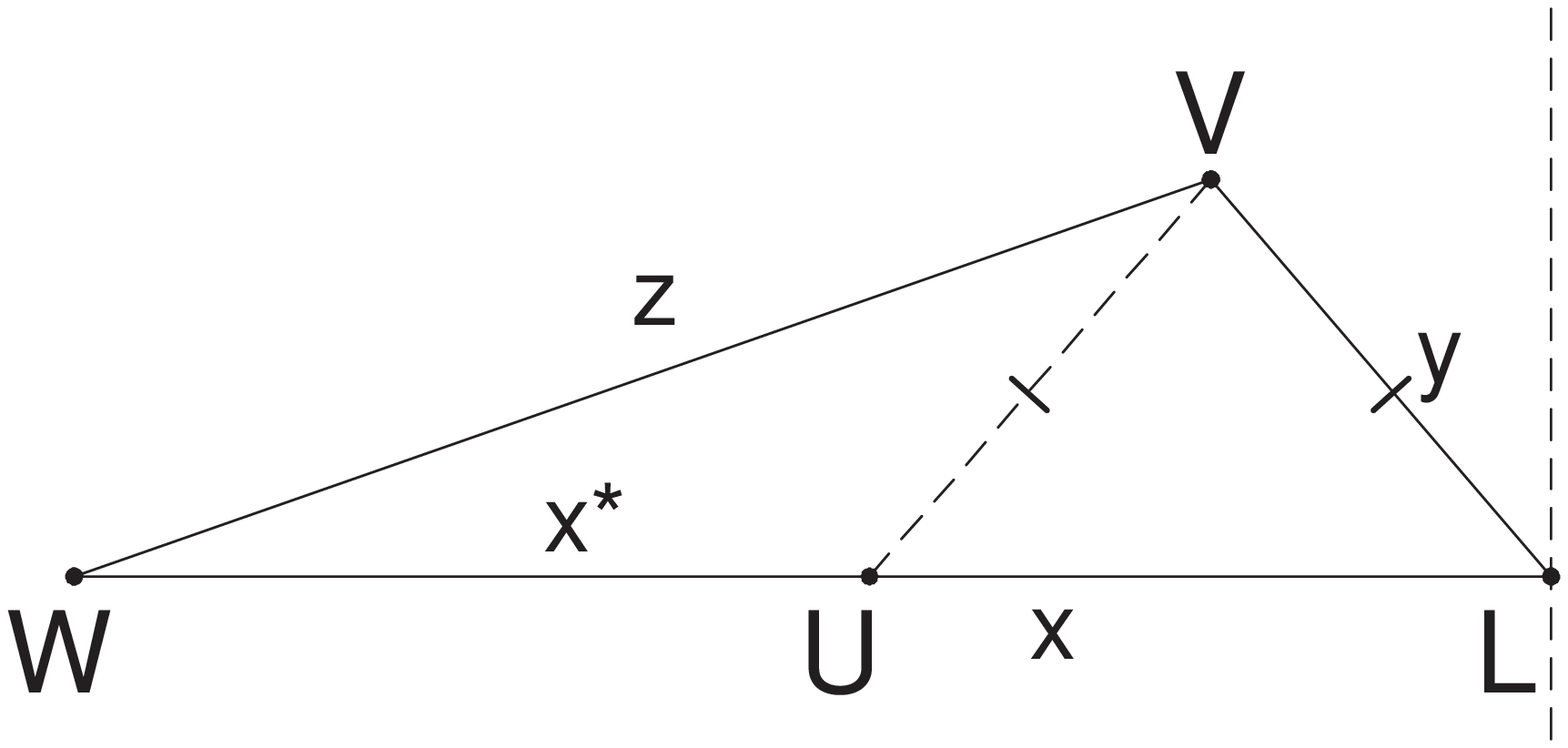,width=3in}}
  \caption{The shortcut triangles.}
  \label{fig:shortcut-triangle}
\end{figure}
Indeed, consider triangle $\triangle WVL$ in Fig.\
\ref{fig:shortcut-triangle}, where $W$ represents source node $w$,
$V$ is destination $v$, and $L$ is $L(v)$. Shortcutting occurs
when destination $V$ is ``in front of'' its landmark $L$, which is
represented by placing $V$ to the left of the vertical line. It is
clear that on small-world graphs, characterized by low average
distances and narrow distance distributions, the possibility for
$V$ to {\em not\/} lie on path $WL$ is minuscule.

To see this, suppose we still want to account for cases when $V$
does not belong to $WL$ by approximating such cases within the
$2$-dimensional Euclidean space. The shortcut path is then $WUV$
in Fig.\ \ref{fig:shortcut-triangle}, with $U$ representing the
first node $u$ on the path from $w$ to $L(v)$ such that $v \in
C(u)$. Its position on $WL$ is defined by $|UV|=|VL|=y$, and the
length of segment $WU$ is denoted by $x^\ast$. The stretch is then
given by $s^\ast = (x^\ast+y)/z$, and solving $\triangle WUV$, we
find that $x^\ast = (z^2-y^2)/x$. Our numerical experiments show
that it does not matter if, with condition $z<x$, we use $s^\ast =
(x^\ast+y)/z$ or $s^\ast = 1$: the two stretch distributions are
virtually identical and the average stretch difference is only in
the fourth digit. This means that $(x^\ast+y)/z$ is virtually
always $1$ as soon as $z<x$. From $(x^\ast+y)/z =
[(z^2-y^2)/x+y]/z = 1$ follows $x=z+y$, which means that $V$ lies
on $WL$.

Note, however, that condition $z<x$ is not the exact condition for
the shortcut presence in the Euclidean space. As mentioned above,
the exact condition is that $V$ is to the left of the vertical
line, which implies $z^2 < x^2 + y^2$. In our experiments, we
observe that using this condition instead of $z<x$ leads to the
stretch distribution drastically distinct from the one observed in
simulations. This is because any estimates based on approximation
of the finite metric space in a graph by the Euclidean metric
space are not applicable to small-world graphs. Such estimates are
more applicable to grid-like graphs with wide distance
distributions and average distances growing as power-law functions
of the graph size.

\subsection{Claim \ref{claim:triangle}}\label{sec:proofs:triangle}

Suppose the distance matrix ${\cal D}$ is given. Let us denote by
${\cal D}(\alpha)$ the set of elements in ${\cal D}$ equal to
$\alpha$ ($\alpha$-elements),
\begin{equation}
    {\cal D}(\alpha) = \big\{ \; {\cal D}_{ij}; \; i,j = 1 \ldots n \; \big|
    \; {\cal D}_{ij} = \alpha \; \big\}.
\end{equation}
Note that the distance p.d.f.\ is equal to the distribution of
values of elements in ${\cal D}$,
\begin{equation}
    f(\alpha) = \frac{|{\cal D}(\alpha)|}{|{\cal D}|}.
\end{equation}
We can also define the set of elements satisfying the triangle
inequality,
\begin{equation}\label{Dabc-def}
    {\cal D}(\beta,\gamma) =
    \big\{ \; {\cal D}_{ij}; \; i,j = 1 \ldots n \; \big| \;
    |\beta-\gamma| \leq {\cal D}_{ij} \leq \beta+\gamma \; \big\} =
    \bigcup_{\alpha = |\beta-\gamma|}^{\beta+\gamma} {\cal D}(\alpha).
\end{equation}
The distribution of $\alpha$-elements in ${\cal D}(\beta, \gamma)$
is
\begin{equation}\label{Dabc-distrib}
    p(\alpha|\beta,\gamma) = \frac{\big| {\cal D}(\alpha) \big|}
    {\big| {\cal D}(\beta,\gamma) \big|} =
    \frac{f(\alpha)}{\sum_{\delta=|\beta-\gamma|}^{\beta+\gamma}
    f(\delta)}
\end{equation}

Suppose that a group of $k$ nodes is randomly selected in the
graph, and that their indices are $\mathbf{i}_k =
\big\{i_1,i_2,\ldots,i_k\big\}$. The distribution of distances
between them is also the distribution of values of elements in a
$k \times k$ submatrix ${\cal D}_{\mathbf{i}_k}$ obtained from
${\cal D}$ by intersecting the $\mathbf{i}_k$ rows and columns.
Since the selection is random, this distribution is also
$f(\alpha)$. If {\em two\/} groups of random nodes,
$\mathbf{i}_{k_1}$ and $\mathbf{j}_{k_2}$, are selected and we are
looking for the distribution of distances between pairs of nodes
belonging to the {\em different\/} groups, we construct a set
${\cal D}_{\mathbf{i}_{k_1},\mathbf{j}_{k_2}}$ by intersecting the
$\mathbf{i}_{k_1}$ rows of ${\cal D}$ with the $\mathbf{j}_{k_2}$
columns and {\em vice versa}. This subset of ${\cal D}$ is no
longer a submatrix of ${\cal D}$, but the distribution of
$\alpha$-elements in it is still $f(\alpha)$.

Our problem is to find the conditional distribution $t(z|x,y)$ for
distance $z$ between a group of $g(x)n$ (on average) random nodes
at distance $x$ from some random node (the landmark closest to the
destination) and another group of $g_1(y)n$ (on average) nodes at
distance $y$ from the same node. The groups may overlap when
$x=y$. They define the subset ${\cal
D}_{\mathbf{i}_{g(x)n},\mathbf{j}_{g_1(y)n}}$ as in the previous
paragraph. The distribution we are looking for, $t(z|x,y)$, is the
distribution of values of elements in this subset. Since there are
no correlations in the distance matrix, the only difference
between this subset and truly random subsets ${\cal
D}_{\mathbf{i}_{k_1},\mathbf{j}_{k_2}}$ from the previous
paragraph is due to the triangle inequality caused by the fact
that the two groups of nodes are neighbors of the same node. This
means that ${\cal D}_{\mathbf{i}_{g(x)n},\mathbf{j}_{g_1(y)n}}$ is
also a subset of ${\cal D}(x,y)$ defined in (\ref{Dabc-def}), and,
since ${\cal D}_{\mathbf{i}_{g(x)n},\mathbf{j}_{g_1(y)n}}$ is
random in other respects, the distribution of values of its
elements is the same as in ${\cal D}(x,y)$. In other words,
$t(z|x,y) = p(z|x,y)$ in (\ref{Dabc-distrib}).

Noticing that, by the formula for the conditional probability,
$t(x,y,z) = t(z|x,y)g(x)g_1(y)$ completes the proof.

\section{The MSR analysis}\label{sec:minimum_analysis}

In this appendix we demonstrate that linearity of functions
$\sigma_{\overline{d},\sigma}^\ast(\overline{d}^\ast)$ in the MSR
follows from the Gaussian form of the distance distribution.

We consider the continuous case, for which
$\overline{s}(\overline{d},\sigma)$ is given by
(\ref{stretch-superlong}). The necessary condition for a local
minimum of $\overline{s}$ in the $\overline{d}$- or
$\sigma$-direction is $\partial \overline{s}/\partial \overline{d}
= 0$ or $\partial \overline{s}/\partial \sigma = 0$ respectively.
After some algebra using
$\erf'(\alpha)=2e^{-\alpha^2}/\sqrt{\pi}$, we obtain, with $x$,
$y$, and $z$ having the same semantics as in
Section~\ref{sec:analytical_results},
\begin{eqnarray}
  \frac{\partial\overline{s}}{\partial\overline{d}}=0
    &\Rightarrow&
        \frac{(x-\overline{d}) + (y-\overline{d}) + (z-\overline{d})}{\sigma}
    + \sqrt{\frac{2}{\pi}}
    \left\{
            \frac{
                e^{-\frac{1}{2}
                    \left( \frac{x+y-\overline{d}}{\sigma}
                    \right)^2}
                - e^{-\frac{1}{2}
                    \left( \frac{|x-y|-\overline{d}}{\sigma}
                    \right)^2}
            }{
                \erf \left( \frac{x+y-\overline{d}}{\sigma\sqrt{2}}
                    \right)
                - \erf \left( \frac{|x-y|-\overline{d}}{\sigma\sqrt{2}}
                    \right)
            }
    \right.
    \nonumber \\
    && \left. {} +
        (q-1) \frac{
                    e^{ -\frac{1}{2}
                        \left( \frac{y-\overline{d}}{\sigma}
                        \right)^2}
                }{
                    1 - \erf \left(
                            \frac{y-\overline{d}}{\sigma\sqrt{2}}
                            \right)
                }
    \right\} = 0, \label{ds-dd-long} \\
    \frac{\partial\overline{s}}{\partial\sigma}=0
    &\Rightarrow&
        \frac{(x-\overline{d})^2 + (y-\overline{d})^2 + (z-\overline{d})^2}{\sigma}
     + \sqrt{\frac{2}{\pi}}
    \left\{
            \frac{
                (x+y-\overline{d}) e^{-\frac{1}{2}
                    \left( \frac{x+y-\overline{d}}{\sigma}
                    \right)^2}
                - (|x-y|-\overline{d}) e^{-\frac{1}{2}
                    \left( \frac{|x-y|-\overline{d}}{\sigma}
                    \right)^2}
            }{
                \erf \left( \frac{x+y-\overline{d}}{\sigma\sqrt{2}}
                    \right)
                - \erf \left( \frac{|x-y|-\overline{d}}{\sigma\sqrt{2}}
                    \right)
            }
    \right.
    \nonumber \\
    && \left. {} +
        (q-1) \frac{
                    (y-\overline{d}) e^{ -\frac{1}{2}
                        \left( \frac{y-\overline{d}}{\sigma}
                        \right)^2}
                }{
                    1 - \erf \left(
                            \frac{y-\overline{d}}{\sigma\sqrt{2}}
                            \right)
                }
    \right\} = 0. \label{ds-ds-long}
\end{eqnarray}
These can be significantly simplified by approximating variables
$x$, $y$, and $z$ by their means (the ``mean field''
approximation): $x \sim \overline{x}$, $y \sim \overline{y}$, and
$z \sim \overline{z}$. Note that $\overline{x} = \overline{z} =
\overline{d}$ and $\overline{y} \leq \overline{d} \Rightarrow
|x-y| \sim \overline{d}-\overline{y}$. Introducing a new variable
\begin{equation}\label{xi-def}
    \xi = \frac{\overline{d}-\overline{y}}{\sigma\sqrt{2}},
\end{equation}
where $\overline{y}$ is, in fact, a function of $\overline{d}$ and
$\sigma$, $\overline{y}\equiv\overline{y}(\overline{d},\sigma)$,
we can reduce (\ref{ds-dd-long}) and (\ref{ds-ds-long}) to,
respectively,
\begin{eqnarray}
  \xi - \frac{q-1}{\sqrt{\pi}} \cdot \frac{e^{-\xi^2}}{1+\erf(\xi)} &=& 0,
  \label{ds-dd-short} \\
   2\sigma\xi\left\{ \xi
   - \frac{q-1}{\sqrt{\pi}} \cdot
        \frac{e^{-\xi^2}}{1+\erf(\xi)}\right\}
   + \sqrt{\frac{2}{\pi}} \cdot
        \frac{\overline{y}e^{-\frac{1}{2}
            \left( \frac{\overline{y}}{\sigma} \right)^2}}
        {\erf\left(\frac{\overline{y}}{\sigma\sqrt{2}}\right)}
        &=& 0. \label{ds-ds-medium}
\end{eqnarray}

We can now search for solutions $\xi_{\overline{d}}^\ast$ of
(\ref{ds-dd-short}) numerically. For $n=10^4$,
$q=[(n/\log_2n)^{1/2}]=27$, which gives a unique solution
$\xi_{\overline{d}}^\ast = 1.32$.

The direct solution of (\ref{ds-ds-medium}) would involve
resolving function $\overline{y}(\overline{d},\sigma)$ first.
However, a simpler way is to use an asymptotic form of the error
function in the last term of (\ref{ds-ds-medium}),
\begin{equation}\label{erf-asympt}
    \erf(\alpha) \sim \frac{2}{\sqrt{\pi}} \, \alpha e^{-\alpha^2},
    \quad \alpha \ll 1,
\end{equation}
which is valid for sufficiently large $\sigma$, $\sigma \gg
\overline{y}$. This reduces (\ref{ds-ds-medium}) to
\begin{equation}\label{ds-ds-short}
    2\xi \left[
    \xi - \frac{q-1}{\sqrt{\pi}} \cdot \frac{e^{-\xi^2}}{1+\erf(\xi)}
    \right] + 1 = 0,
\end{equation}
which has a unique solution $\xi_\sigma^\ast = 1.24$.

To see that $\sigma_{\overline{d},\sigma}^\ast(\overline{d}^\ast)$
are linear, we just need to check that
$\overline{y}(\overline{d},\sigma)$ is a linear function of its
arguments in the MSR. Indeed,
\begin{equation}
    \overline{y}(\overline{d},\sigma) = \int y g_1(y) \, \dif y =
    \frac{q}{\sigma\sqrt{2\pi}2^{q-1}}
    \int y
    e^{-\frac{1}{2}\left(\frac{y-\overline{d}}{\sigma}\right)^2}
    \left[1-\erf\left(\frac{y-\overline{d}}{\sigma\sqrt{2}}\right)\right]^{q-1}
    \dif y.
\end{equation}
Changing variables, $\zeta=(y-\overline{d})/(\sigma\sqrt{2})$, and
using the asymptotic form of the error function,
(\ref{erf-asympt}), we see that
\begin{equation}
    \begin{array}{cc}
      \overline{y}(\overline{d},\sigma) \sim c_1\sigma +
      c_2\overline{d},
      \quad \text{where} &
      \begin{array}{rcl}
        c_1 &=& \sqrt{\frac{2}{\pi}} \cdot \frac{q}{2^{q-1}}
        \int \zeta e^{-\zeta^2} \left(1-\frac{2}{\sqrt{\pi}}\zeta
        e^{-\zeta^2}\right)^{q-1}\dif\zeta \\
        c_2 &=& \frac{1}{\sqrt{\pi}} \cdot \frac{q}{2^{q-1}}
        \int e^{-\zeta^2} \left(1-\frac{2}{\sqrt{\pi}}\zeta
        e^{-\zeta^2}\right)^{q-1}\dif\zeta
      \end{array}
    \end{array}.
\end{equation}
Substituting this into (\ref{xi-def}), we find that
\begin{equation}
    \sigma_{\overline{d},\sigma}^\ast(\overline{d}^\ast) \sim
    c_{\overline{d},\sigma}\overline{d}^\ast =
    \frac{1-c_2}{\xi_{\overline{d},\sigma}^\ast\sqrt{2}+c_1}
    \,\overline{d}^\ast.
\end{equation}
Numerical evaluations of $c_{1,2}$ yield $c_{\overline{d}} = 0.53$
and $c_\sigma = 0.57$. We have a good match between
$c_{\overline{d}}$ and the data fit (cf.\
Fig.~\ref{fig:minimum}(d)), $0.57$. The match between $c_\sigma$
and the data fit, $0.79$, is worse, which suggests that the source
of error is mostly in (\ref{erf-asympt}). We do not obtain
non-zero values for the additive coefficients,
$\tilde{c}_{\overline{d},\sigma}$, in
$\sigma_{\overline{d},\sigma}^\ast(\overline{d}^\ast) \sim
c_{\overline{d},\sigma}\overline{d}^\ast +
\tilde{c}_{\overline{d},\sigma}$ that define the apex location.
Thus, we conclude that a more accurate analysis of the essentially
discrete case with small $\sigma$ is required to analytically
obtain the apex location.

Note, however, that equations (\ref{ds-dd-short}) and
(\ref{ds-ds-medium}) are consistent with the observed analytical
structure of $\overline{s}(\overline{d},\sigma)$ even for small
$\sigma$. Indeed, the solutions of the {\em system\/} of equations
(\ref{ds-dd-short}) and (\ref{ds-ds-medium}) (corresponding to the
true stationary points of  $\overline{s}$, $\dif \overline{s}=0$)
exist only for $\sigma\to0$, when the last term of
(\ref{ds-ds-medium}) goes away. Then we have from (\ref{xi-def})
with $\xi=\xi_{\overline{d}}^\ast$ that
$\overline{y}\to\overline{d}$ as expected, and {\em any\/}
$\overline{d}$ delivers a solution. As discussed in
Section~\ref{sec:minimum}, the actual average distance can be only
an integer $k$ or $k+1/2$ when $\sigma\to0$. Thus, the flat-topped
peaks and narrow cracks we observe in Fig.~\ref{fig:minimum}(c) at
$\sigma=0$ are consistent with $\dif \overline{s}=0$ there.

Since $\dif \overline{s} = 0$ only when $\sigma\to0$, we have also
effectively demonstrated, at least with the approximations we have
made to obtain (\ref{ds-dd-short}) and (\ref{ds-ds-medium}), that
the apex {\em cannot\/} be a {\em true\/} stationary point of
$\overline{s}$.

\begin{figure}
\centerline{\epsfig{file=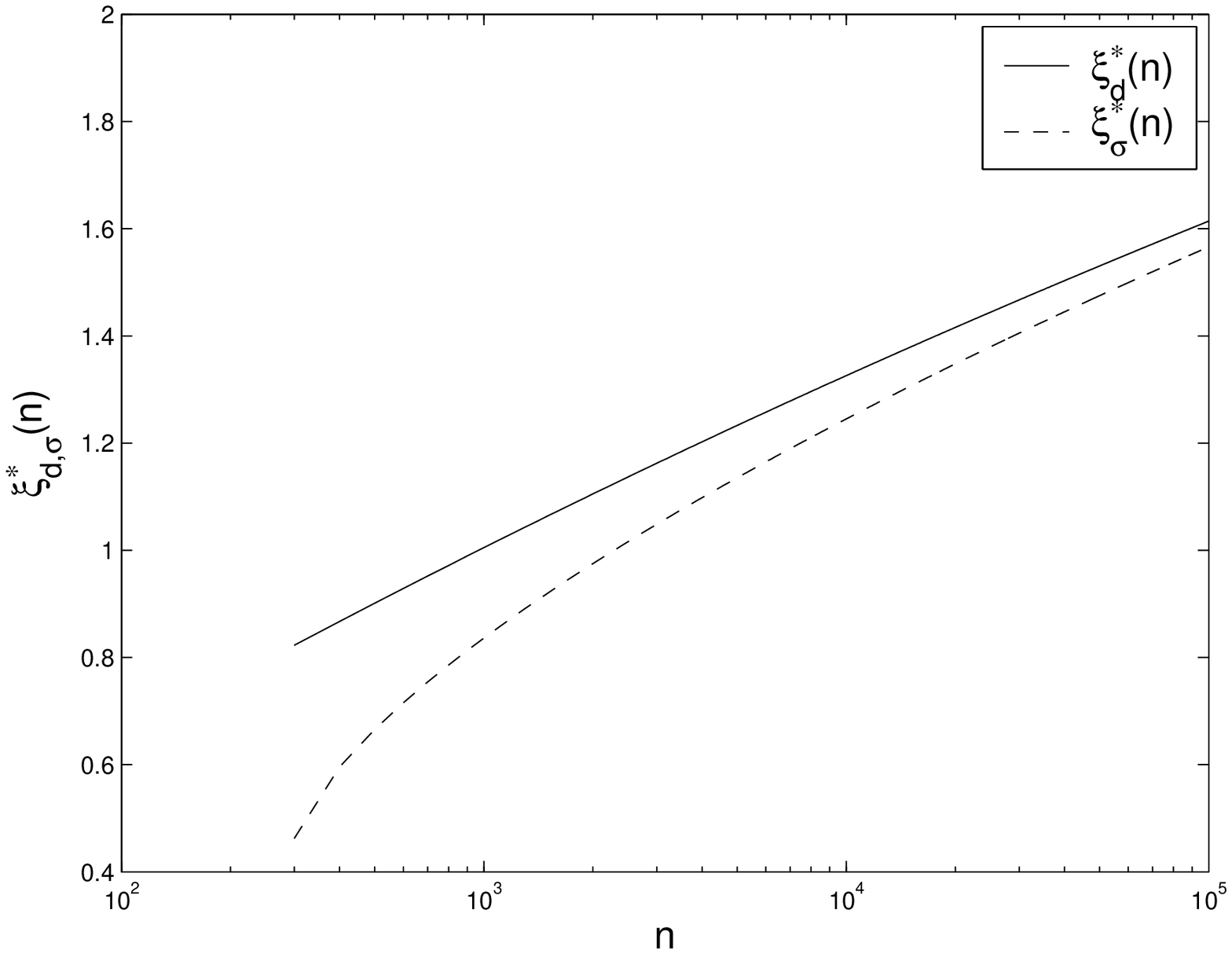,width=3in}}
\caption{Solutions of equations (\ref{ds-dd-short}) and
(\ref{ds-ds-short}), $\xi_{\overline{d}}^\ast$ (solid line) and
$\xi_\sigma^\ast$ (dashed line), as functions of the graph size
$n$.}\label{fig:xistar}
\end{figure}

The LS size $q$ is a function of $n$, and, hence, solutions
$\xi_{\overline{d},\sigma}^\ast$ are also functions of the graph
size. They are shown in Fig.~\ref{fig:xistar}. We see that
$\xi_{\overline{d},\sigma}^\ast(n) = \Theta(\log n)$. This sheds
some light on the analytical structure of $\overline{s}$ as a
function of $n$. We can see, for example, that as the network
grows, the MSR becomes narrower and closer to $\sigma=0$. Since
for scale-free nets, $\sigma$ does approach $0$ as $n\to\infty$
(\cite{dorogovtsev-private}), we conclude that, independent of
their size, the scale-free graphs are characterized by the lowest
possible average stretch values.

\end{document}